%% file: main.tex
\documentclass[a4paper,11pt]{article}
\pdfoutput=1

\usepackage[margin=1in]{geometry}
\usepackage{array}
\usepackage{bigstrut}

\usepackage{algorithm}
\usepackage{algpseudocodex}

\usepackage{bbm}
\usepackage[utf8]{inputenc}
\usepackage[english]{babel}
\usepackage[T1]{fontenc}
\usepackage{amsmath,amssymb,amsthm}
\usepackage{mathtools}%

\usepackage[font=small,labelfont=bf]{caption}
\usepackage{float}
\usepackage{tikz}
\usetikzlibrary{positioning,calc,patterns,arrows,decorations.pathreplacing}
\definecolor{pmint}{HTML}{C4E8D1}     %
\definecolor{pblue}{HTML}{C9DDF2}     %
\definecolor{pbluedeep}{HTML}{93BCE3} %
\definecolor{plav}{HTML}{DDCDF0}      %
\definecolor{prose}{HTML}{F6C1CA}     %
\definecolor{prosepale}{HTML}{FBE3E7} %
\definecolor{ppeach}{HTML}{FFDDBE}    %
\definecolor{ppeachdeep}{HTML}{FBC08F}%
\definecolor{pgray}{HTML}{F2F0ED}     %
\definecolor{pline}{HTML}{8E95A3}
\definecolor{trose}{HTML}{B0576B}
\definecolor{tlav}{HTML}{7A5EA8}
\definecolor{tblue}{HTML}{4472A3}
\definecolor{tgreen}{HTML}{3F8A5F}
\definecolor{tpeach}{HTML}{B9712F}
\colorlet{varfill}{pmint}
\colorlet{chkfill}{pblue}
\colorlet{msgcolor}{tlav}
\tikzset{
  var/.style       = {circle, draw=black, fill=pmint, minimum size=2.6mm, inner sep=0pt},
  chk/.style       = {rectangle, draw=black, fill=pblue, minimum size=2.4mm, inner sep=0pt},
  gennode/.style   = {rectangle, draw=black, fill=plav, minimum size=2.8mm, inner sep=0pt},
  erased/.style    = {fill=prose},
  erasedpale/.style= {fill=prosepale},
  known/.style     = {fill=white},
  unsat/.style     = {fill=pbluedeep},
  tedge/.style     = {pline, thin},
  ghost/.style     = {draw=black!20, fill=black!5},
  ghostedge/.style = {black!15, thin},
  ghosttext/.style = {text=black!35},
  lbl/.style       = {font=\small, inner sep=2pt},
  msg/.style       = {tlav, thick, -{Latex[length=2mm]}},
  msglbl/.style    = {text=tlav, font=\small\bfseries},
  setbox/.style    = {draw=trose, thick, rounded corners=2pt, inner sep=3pt},
  setlbl/.style    = {text=trose},
  channel/.style   = {-{Latex[length=2mm]}, decorate,
                      decoration={snake, amplitude=0.6mm, segment length=2.5mm, post length=2.5mm}},
  pbrace/.style    = {decorate, decoration={brace, amplitude=4pt}, pline, thick},
  quad/.style      = {draw=pline, rounded corners=4pt, thick},
  gridline/.style  = {draw=white, line width=0.8pt},
  arrowto/.style   = {-{Latex[length=1.8mm]}, thick},
}

\usepackage{setspace}
\usepackage[export]{adjustbox}
\usepackage{makecell}
\usepackage[inline]{enumitem}
\usepackage{url}

\usepackage{xcolor}
\definecolor{darkblue}{RGB}{0,0,128}
\definecolor{darkgreen}{RGB}{0,150,0}

\usepackage[numbers,sort&compress]{natbib}
\usepackage[pdfusetitle]{hyperref}
\hypersetup{breaklinks, colorlinks, linkcolor=blue, citecolor=magenta, filecolor=red, urlcolor=darkblue}

\usepackage{thmtools}
\usepackage{thm-restate}

\newtheorem{theorem}{Theorem}
\newtheorem*{theorem*}{Theorem}
\newtheorem{lemma}[theorem]{Lemma}

\newtheorem{claim}[theorem]{Claim}
\newtheorem{corollary}[theorem]{Corollary}
\newtheorem{fact}[theorem]{Fact}
\newtheorem{definition}[theorem]{Definition}

\newtheorem{obs}[theorem]{Observation}

\usepackage{mystyle}
\graphicspath{{../}{./}{src_tikz/}}

\usepackage{authblk}

\title{Linear-time erasure recovery for quantum expander codes}
\author[1]{Anirudh Krishna}
\author[2]{Christopher A. Pattison}
\affil[1]{ {\it {\small IBM Quantum, T. J. Watson Research Center, Yorktown Heights, New York 10598, USA}}}
\affil[2]{ {\it {\small Simons Institute for the Theory of Computing, University of California, Berkeley, California 94720, USA}}}
\date{}

\begin{document}
\maketitle

\begin{abstract}
    Erasures are used to model faults at a subset of known locations in various platforms used to construct fault-tolerant quantum computers.
    Beyond the utility in erasure recovery, efficient decoders for erasures are useful to both understand error decoders and as a subroutine for error decoding.
    Classical expander codes, which belong to the family of low-density parity-check (LDPC) codes, can be equipped with the Peeling decoder which works in linear time. 
    Our main result is the algorithm \emph{Small-Set-Peeling}, a quantum generalization of the classical Peeling decoder for erasures, that applies to quantum expander codes (Tillich, Z\'emor, IEEE T-IT 2013; Leverrier, Tillich, Z\'emor, FOCS 2015).
    This decoder can correct up to $D-1$ erasures, where $D$ is the (design) distance of the code and, importantly, is the first decoder that is guaranteed to do so in linear time.
    Combining with the erasure-conversion algorithm Small-Set-Find (Krishna, Livni-Navon, Wootters, SODA 2024), our result also implies the existence of a novel graph-local linear-time decoder for Pauli errors.
\end{abstract}
    
\input{01-intro}
\input{02-technical-overview}

\pagebreak
\input{03-background}
\input{04-erasure_decoding}

\pagebreak
\bibliographystyle{alpha}
\bibliography{references}

\appendix
\end{document}

%% file: 01-intro.tex
\section{Introduction}

An \emph{erasure} is a model for the loss of information during storage and transmission.
On $n$ bits, an erasure is abstractly represented by a set $L \subseteq [n]$ which is interpreted to mean that the bits in $L$ are just as likely to be flipped as not, and therefore completely unreliable; bits outside $L$ are completely unaffected.
As $L$ itself is known in this model, recovering from erasures is easier than recovering from errors whose location is unknown: In worst-case settings, a (binary linear) code $\ccode \subseteq \bbF_2^{n}$ that is robust to at most $d-1$ erasures is only robust to $\lfloor (d-1)/2\rfloor$ errors.
For real-world implementation, we can equip a code with an efficient decoder, an algorithm to repair corrupted codewords.
Expander codes are a class of constant-rate low-density parity-check (LDPC) codes that can be equipped with the \peeling decoder that can correct a constant-fraction of erasures and only requires $O(n)$ time.

Erasures are also used to model the loss of \emph{quantum} information.
One way erasures manifest is as depolarizing noise on known locations \cite{grassl1997codes}.
Similar to the classical erasure, this is interpreted to mean that qubits in some (known) locations are completely unreliable.
In photonic architectures, erasures are used to model lost photons which manifest as a missing detection \cite{bartolucci2023fusion}.
Erasures in the quantum setting are easier to correct than the more general case of Pauli errors on unknown locations: Erasures on an $N$-qubit quantum code can be decoded in $O(N^3)$ time.
Motivated by this observation, several qubit platforms, including neutral atoms~\cite{wu2022erasure}, superconducting circuits~\cite{kubica2023erasure} and trapped ions~\cite{kang2023quantum}, can be engineered so that erasures are their dominant failure mode: The qubit leaves its computational subspace, and this event can be detected without disturbing the encoded state.
At the same time, quantum LDPC codes, are being explored as avenues for efficient quantum error correction in practice~\cite{xu2024constant,bravyi2024high}.
Constant-rate \emph{quantum} expander codes, a family of $\dsl N,K,D\dsr$ quantum LDPC codes where $K = \Theta(N)$ and distance $D = \Theta(\sqrt{N})$, serve as a useful theoretical testbed for more practically relevant code families where progress is naturally more numerical in nature~\cite{kovalev2013quantum,panteleev_degenerate_2021,bravyi2024high,lee2026design}.
To this end, the complexity of the decoder is crucial for enabling real-time decoding \cite{terhal2015quantum}, and we would like to do better than the generic cubic time $O(N^3)$ time algorithm.
With this motivation, and inspired by the classical setting, it is natural to ask whether we can equip quantum expander codes with linear-time decoders for erasures.

\textbf{Our main result is a linear-time erasure decoder for constant-rate quantum expander codes.}
We propose and analyze a decoder called $\sspeeling$ for quantum expander codes which generalizes the classical $\peeling$ decoder.
We highlight some implications of our result.

\begin{enumerate}
    \item \textbf{A novel decoder for erasures.} Our result represents the first provable linear-time algorithm for correcting erasures on constant-rate quantum LDPC codes that can correct erasures of any size less than the \emph{design} distance $D$.\footnote{Our result is stated in terms of the design distance rather than the code distance, as the latter often cannot be directly computed (See Lemma~\ref{lem:distance}). Analyses of $\ssflip$ are also measured against the design distance, so this does not affect our comparisons.}
    This stands in contrast to the use of a linear-time error decoding algorithm such as \(\ssflip\)~\cite{leverrier2015quantum,fawzi2018efficient,fawzi2020constant} as an erasure decoding algorithm: A decoding algorithm correcting \(t\) errors can correct \(t\) erasures, but such algorithms may only guarantee the ability to correct a number of errors less than half of the distance \(D\).
    In contrast, we can correct all erasures of size \emph{up to $D-1$}.

    \item \textbf{A linear-time decoder for Pauli errors.}
    The fact that recovery from erasures is easier than recovery from errors whose location is unknown also motivates a way to design decoders for errors: Given an unknown error $\err$, suppose we can efficiently find an envelope $\env \subseteq \qubits$ such that $\err \subseteq \env$.
    Error correction then becomes a two-stage process: First, use an envelope-finding algorithm, and then implement an erasure-recovery algorithm.
    Most famous among these is the \unionfind decoder \cite{delfosse2021almost, delfosse2021union, delfosse2022toward} which works in (almost) linear time.
    The first work to design and study an envelope-finding algorithm for \emph{constant-rate} quantum LDPC codes was \cite{krishna2024viderman} which proposed and analyzed an algorithm called $\ssfind$ for quantum expander codes.
    Given the syndrome $\synd(\err)$ of an error $\err$, this algorithm can find an envelope $\env$ that is guaranteed to contain $\err$ in $O(N)$ time.
    However, an $O(N)$-time decoder for erasures was not known at the time.
    Combining with our result implies the existence of a graph-local linear-time decoder $\ssfind+\sspeeling$ for correcting errors.\footnote{See also the recent work \cite{ostrev2026find} discussed in related work.}
    \item \textbf{Low-complexity and graph-locality.} Our algorithm is an iterative, graph-local algorithm: In each iteration, it makes an elementary decision which only involves a small (constant) number of qubits.
    Graph-local processing is frequently a useful property for constructing and analyzing decoders in the stochastic-noise setting~\cite{fawzi2018efficient}.
    In particular, our decoder retains its linear runtime under stochastic noise.\footnote{Each connected component of the erasure may be corrected independently by invoking the algorithm on each connected component with the syndrome outside of the erasure set to zero. Since the number of iterations scales as the component size, the overall runtime is linear in the block length.}
    \item \textbf{Parameters for guarantee.} Our guarantee for correcting erasures with $\sspeeling$ holds under a weaker expansion assumption than that needed for $\ssflip$.
    Note, however, that the expansion requirement for $\ssfind+\sspeeling$ that corrects Pauli errors is determined by $\ssfind$ whose expansion requirements are more demanding than $\ssflip$. Oftentimes, decoders for quantum expander codes are later shown to work well on practically relevant code families, and we expect this to be the case for our algorithm.
\end{enumerate}

With the motivation for our work in hand, we now describe it in further detail.

\subsection{Our Result}
We introduce some notation to state our results (see Section~\ref{sec:background} for full definitions).
An $\dsl N,K, D \dsr$ \emph{quantum error correcting code} $\qcode$ encodes a set of $K$ logical qubits in a set $\qubits$ of $N$ qubits.
The \emph{distance} $D$ of $\qcode$ (formally defined in Section~\ref{sec:background}) is the number of erasures that the code can correct: $\qcode$ is robust to all erasures of size $D-1$.

The codes we consider can be specified by two sets $\pcs$ and $\gens$, where each element of $\pcs$ or $\gens$ is a quantum generalization of a parity check and specifies a constraint on the qubits.
Specifically, we consider \emph{quantum expander codes}, a family of constant-rate quantum LDPC codes \cite{tillich2013quantum, leverrier2015quantum}.
Quantum expander codes are graph-based codes, i.e.\ they can be described by a graph $\cG$ whose vertices are the qubits $\qubits$ and the sets $\pcs$ and $\gens$; the constraints imposed by elements of $\pcs$ and $\gens$ correspond to the edges of $\cG$.
A quantum code is a \emph{quantum LDPC code} if the graph $\cG$ is \emph{sparse}.

For Calderbank-Shor-Steane (CSS) quantum codes, it suffices to consider only two types of errors: \emph{$X$-type errors} and \emph{$Z$-type errors}
The constraints $\pcs$ are meant to correct $Z$-type errors, while the constraints $\gens$ are meant to correct $X$-type errors.
It suffices to deal with these two types individually.
For the rest of this paper, we focus on $Z$-type errors and mention that $X$-type errors can be corrected in an identical manner.
With this understanding, we refer to $\pcs$ as ``parity-checks,'' as they are used to detect the errors and $\gens$ simply as ``generators''.  

An erasure is specified by a subset $\env \subseteq \qubits$.
For sufficiently small $\env$, we would like a decoder with the following property.
For all ($Z$-type) errors $\err$ supported within $\env$, the decoder takes the syndrome $\synd(\err)$ and $\env$ as input and identifies the error $\err$.
The maximum size of $\env$ that the decoder can tolerate is called the \emph{decoding radius}.  

\paragraph{A decoder for erasures.}
Our main result is the first provable linear-time decoder for erasures \emph{Small-Set-Peeling} ($\sspeeling$) on constant-rate quantum LDPC codes whose decoding radius is the design distance.

\begin{restatable}{theorem}{maininformal}
\label{thm:maininformal}
    Let $\qcode$ be a $\dsl N,K, D\dsr$ quantum expander code (See Section~\ref{sec:background} for a formal definition) obtained from a bipartite, $(\dV,\dC)$-biregular, bidirectional expander graph $G$ on vertex set $V \sqcup C$, where $V = [n]$, $C = [m]$ and expansion parameter $\eps < 1/4$.
    There exists a decoder $\sspeeling$ such that, for all $\env \subseteq \qubits$ such that $\wt{\env} < D$, we have the following guarantee:
    For any ($Z$-type) error $\err \subseteq \env$, $\sspeeling$ takes as input the syndrome $\synd(\err)$ and the envelope $\env$ as input, and returns a correction $\corr$ (up to an equivalence class; See Section~\ref{sec:background}) in $O_{\dV,\dC}(\wt{\env})$ time.
\end{restatable}

The decoder takes $O(\wt{\env})$ time and thus linear in the size of the erasure.
We highlight also that it is sufficient to have $\eps < \frac{1}{4}$, which is weaker than the demand on decoders: \cite{leverrier2015quantum} requires $\eps < \frac{1}{6}$.

We remark that the critical new ingredient compared to prior work is the exploitation of \emph{degeneracy} in a graph-local algorithm:
We loosen the erasure decoding problem to permit the correction to be supported outside of the erasure as long as the returned correction is equivalent to the erasure (i.e. the residual error after correction is trivial).

\paragraph{A decoder for errors.}
Given a (reduced) error $\err \subseteq \qubits$, an erasure-conversion algorithm takes the syndrome $\synd(\err)$ as input and returns an envelope $\env$ such that $\err \subseteq \env$.
The work \cite{krishna2024viderman} presented an algorithm $\ssfind$, a generalization of a classical algorithm designed by Viderman for classical expander codes \cite{viderman2013linear}.
We point the interested reader to \cite{krishna2024viderman} for a formal description of $\ssfind$.
By combining $\ssfind$ with an erasure decoder, it can be used to correct errors.
The algorithm $\ssfind$ is itself a linear-time algorithm, and therefore, by combining with a linear-time erasure decoder, this yields a new graph-local linear-time decoder for errors that is distinct from $\ssflip$.

\begin{corollary}
\label{cor:error-decoder}
    Let $\mathcal{H}$ be an $\dsl N,K,D \dsr$ quantum expander code on qubits $\cQ$ constructed from an expander graph $G$ with left-degree $\Delta_V$ and right-degree $\Delta_C$, and expansion parameter $\epsilon < 1/10$.

    Let $\err$ be a (reduced) error pattern with weight at most 
    \[ |\err| < \gamma \cdot D - \Delta_V, \]
    where 
    $ \gamma = \frac{\Delta_V}{\Delta_C} \cdot \frac{1 - 10 \epsilon}{4}$, and let $\synd(\err) \subseteq \pcs$ be the syndrome of $\err$.
    Then $\ssfind + \sspeeling$, given the syndrome $\synd(\err)$, identifies $\err$ in time $O_{\Delta_V, \Delta_C}(|\err|)$.
\end{corollary}

\subsection{Related work}

\paragraph{Erasure recovery.}
For surface codes, Delfosse and Z\'emor~\cite{delfosse2020linear} gave a linear-time maximum-likelihood erasure decoder.
This then forms the second stage of the $\unionfind$ decoder \cite{delfosse2021almost} which is known to work in (almost) linear time.
While surface codes are LDPC codes, with each qubit only needing to connect to 4 other qubits in its immediate neighborhood, they are only capable of encoding a single logical qubit, irrespective of code size.
Turning to constant-rate codes, Connolly, Londe, Leverrier and Delfosse~\cite{connolly2022fast} applied $\peeling$ to hypergraph product codes and identified stabilizer generators as the obstacle to peeling.
This observation is important for understanding $\sspeeling$ as well and we will return to it in Section~\ref{sec:overview}.
Connolly \emph{et al}.\ overcome this problem using a decoder they called \emph{pruned peeling}, which removes a qubit from the erasure whenever the support of a generator lies entirely inside it, and the \emph{VH decoder}, which splits the remaining erasure into vertical and horizontal clusters, i.e.\ subsets of a single row or column of the product, and corrects them using Gaussian elimination.
Their decoder runs in $O(N^2)$ time and performs close to maximum likelihood in simulations.

Prabhu, Vaishya, Bhatnagar, Kolhe, Lalitha \& Kumar~\cite{prabhu2025efficacy} also study hypergraph product codes with $\peeling$ followed by a cluster-based decoder and $\ssflip$ numerically, and state a worst-case guarantee for a version of $\ssflip$ whose flips are restricted to the erasure.

Pech, G\"okduman, Yao, Pfister~\cite{pech2026stabilizer} generalize pruned peeling with a \emph{dual peeling} step.
Combined with $\peeling$, this is shown to be maximum likelihood decoding for surface codes; for other codes, its performance is studied numerically.
The time complexity is also considered empirically.
Belief propagation with guided decimation has also been applied to erasures by G\"okduman, Yao, Pfister~\cite{gokduman2024erasure}.

More recently, \cite{ostrev2026find} constructed a linear-time decoder using linear-algebraic techniques that achieves a decoding radius \(\frac{1-3 \epsilon}{1-2 \epsilon} D\).
This suffices to establish Corollary~\ref{cor:error-decoder} against adversarial errors.
This is in contrast to our decoder which can correct erasures of size up to $D-1$.
Furthermore, the non-graph-local nature of the algorithm may mean it is difficult to preserve the linear time complexity in a stochastic noise setting.

\paragraph{Linear-time decoders for Pauli errors with provable guarantees.} The \emph{Small-Set-Flip} or $\ssflip$ decoder of Leverrier, Tillich \& Z\'emor \cite{leverrier2015quantum} was the first provable linear-time decoder for adversarial Pauli errors on constant-rate quantum expander codes.
It has been generalized in subsequent works to cover other noise models such as stochastic errors and in scenarios where even the syndrome information is unreliable \cite{fawzi2018efficient,fawzi2020constant}.
$\ssflip$ also forms the basis for linear-time decoders for quantum LDPC codes with both constant rate and relative distance \cite{lin2022good,leverrier2023decoding,gu2022efficient,dinur2022good}.

\subsection{Acknowledgements}
The decoder $\sspeeling$ and its analysis was developed with the help of an LLM, which built on core ideas and direction from the authors.
The tool materially affected Section~\ref{sec:decoder}.
We have verified all results and take responsibility for all claims made in this paper.
CAP is currently a Simons-CIQC postdoctoral fellow at the Simons Institute for the Theory of Computing, supported by NSF QLCI Grant 2016245.

%% file: 02-technical-overview.tex
\section{Technical Overview}
\label{sec:overview}

In this section, we present a more detailed overview of our result.
We begin by discussing expander codes, a type of classical LDPC codes, and the $\peeling$ decoder in Section~\ref{subsec:peeling-overview}.
We introduce quantum expander codes and discuss why $\peeling$ does not directly generalize in Section~\ref{subsec:sspeeling-motivation-overview}.
This then motivates the algorithm Small-Set-Peeling or $\sspeeling$ which we present informally in Algorithm~\ref{alg:sspeeling-informal}.

\subsection{Expander codes \& Peeling}
\label{subsec:peeling-overview}

An $[n,k,d]$ (classical, binary, linear) error correcting code $\ccode$ is a linear subspace of $\mathbb{F}_2^n$. 
A code $\ccode$ of \emph{distance} $d$ is robust to all erasures of size strictly less than $d$: For any erasure $L \subseteq [n]$ such that $\wt{L} < d$, we can uniquely recover a codeword \(x\) of \(\ccode\) given access only to the complement of the erasure \([n] \setminus L\).

A classical LDPC code is defined by a sparse bipartite graph $G$ over vertex set $V \sqcup C$ in the following way.
We associate the (left) vertices $V$ (with $|V| = n$) with the symbols of a codeword, and the (right) vertices $C$ with \emph{parity-checks}. 
We let $\nbhd(c)$ denote the neighbors of $c$ in $G$.

We say that $E \subseteq [n]$ is an error if there exists at least one parity-check $c \in C$ such that $\wt{\nbhd(c) \cap E} \equiv 1 \pmod{2}$.
We focus on the case where the underlying graph $G$ is a bipartite expander (see Definition~\ref{def:exp}); such a code is sometimes called an \emph{expander code}~\cite{sipser1996expander}.

\begin{figure}[ht]
  \centering
  \begin{tikzpicture}[x=0.9cm, y=1.4cm,scale=0.9,
      row/.style={lbl, text height=1.6ex, text depth=0.25ex}]
      \node at (-0.5,1.4) {(\emph{a})};
    \foreach \i in {1,2,6}
      \node[var, known, label={[lbl,yshift=-20, name=v\i-label]:$ $}] (v\i) at (\i-1, 0) {};
    \foreach \i in {3, 4, 5}
      \node[var, erased, label={[lbl,yshift=-20, name=v\i-label]:$ $}] (v\i) at (\i-1, 0) {};
    \node[chk, label={[lbl]above:$c_1$}] (c1) at (1.0, 1) {};
    \node[chk, label={[lbl]above:$c_2$}] (c2) at (2.1, 1) {};
    \node[chk, label={[lbl]above:$c_3$}] (c3) at (3.9, 1) {};
    \foreach \c/\v in {c1/1, c1/2, c1/3, c2/2, c2/4, c3/4, c3/5, c3/6}
      \draw[tedge] (\c) -- (v\v);
    \foreach \i/\b in {1/0, 2/1, 3/1, 4/1, 5/0, 6/1}
      \node[row] at (\i-1, -0.27) {$\b$};
    \draw[purple] (1.75,-0.4) rectangle (4.25,0.2);
    \node[purple] at (3,-0.6) {$L$};
    \begin{scope}[xshift=160]
          \node at (-0.5,1.4) {(\emph{b})};
        \foreach \i/\b in {1/0, 2/1, 6/1}
          \node[var, known, label={[lbl]below:$\b$}] (v\i) at (\i-1, 0) {};
        \foreach \i/\b in {3/0, 4/0}
          \node[var, label={[lbl]below:$\b$}] (v\i) at (\i-1, 0) {};
        \foreach \i/\b in {5/0}
          \node[var, erased, label={[lbl]below:$\b$}] (v\i) at (\i-1, 0) {};
        \node[chk, label={[lbl, name=c1-label]above:$c_1$}] (c1) at (1.0, 1) {};
        \node[chk, label={[lbl, name=c2-label]above:$c_2$}] (c2) at (2.1, 1) {};
        \node[chk, label={[lbl]above:$c_3$}] (c3) at (3.9, 1) {};
        \foreach \c/\v in {c1/1, c1/2, c1/3, c2/2, c2/4, c3/4, c3/5, c3/6}
          \draw[tedge] (\c) -- (v\v);
        \node[setbox, fit=(c1) (c2) (c1-label) (c2-label)] (Sbox) {};
        \node[setlbl, left=1pt of Sbox] {$U$};
    \end{scope}
    \begin{scope}[xshift=320]
          \node at (-0.5,1.4) {(\emph{c})};
        \foreach \i/\b in {1/0, 2/1, 3/1}
          \node[var, ghost, label={[lbl, ghosttext]below:$\b$}] (v\i) at (\i-1, 0) {};
        \foreach \i/\b in {4/1, 6/1}
          \node[var, known, label={[lbl]below:$\b$}] (v\i) at (\i-1, 0) {};
          \foreach \i/\b in {5/0}
          \node[var, erased, label={[lbl]below:$\b$}] (v\i) at (\i-1, 0) {};
        \node[chk, label={[lbl]above:$c_1$}, ghost] (c1) at (1.0, 1) {};
        \node[chk, label={[lbl]above:$c_2$}, ghost] (c2) at (2.1, 1) {};
        \node[chk, label={[lbl, name=c3-label]above:$c_3$}] (c3) at (3.9, 1) {};
        \foreach \c/\v in {c1/1, c1/2, c1/3, c2/2, c2/4}
          \draw[ghostedge] (\c) -- (v\v);
        \foreach \c/\v in {c3/4, c3/5, c3/6}
          \draw[tedge] (\c) -- (v\v);
        \draw[purple] (3.75,-0.4) rectangle (4.25,0.2);
        \node[purple] at (4,-0.6) {$L$};
    \end{scope}
  \end{tikzpicture}
  \caption{Illustrating $\peeling$. ($a$) is the Tanner graph of the code, together with a valid codeword $(0,1,1,1,0,1)$.
  $L$ indicates the erasure location.
  For this example, all bits in $L$ are replaced by $0$.
  ($b$) shows $U$, the checks $c_1$ and $c_2$ which are only incident to a single location in $L$.
  The variable nodes in green will be corrected because they have unique neighbors in $U$.
  ($c$) shows the code after $c_1$ and $c_2$ are peeled.
  Peeled portions of the Tanner graph are grayed out.}
  \label{fig:peeling-channel}
\end{figure}
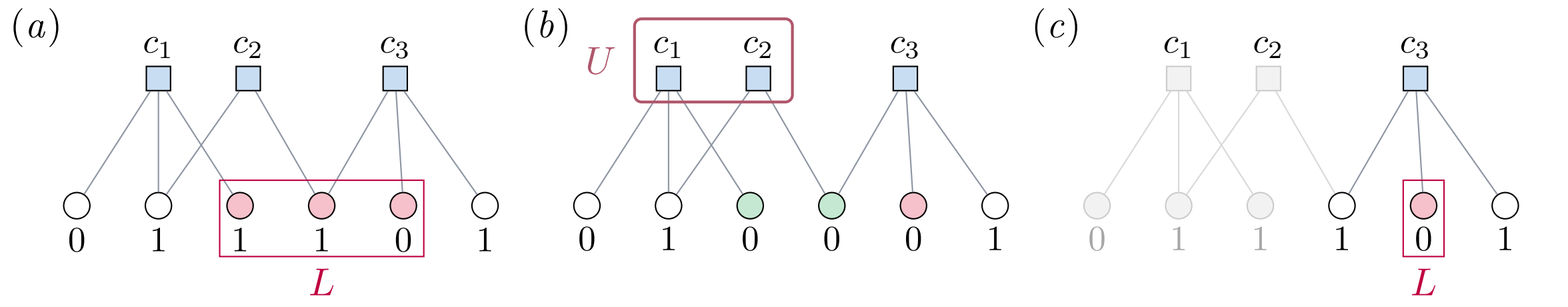

The decoding problem is described as follows: Consider an erasure $L \subseteq [n]$.
We say that we can correct the erasure if, for all errors $E \subseteq L$, we can design a decoder that accepts as input the syndrome $\synd(E)$ and the erasure $L$, and finds $E$.
For classical expander codes, erasures can be corrected using a linear-time decoder called $\peeling$ that we describe in this section.
The decoder uses the following elementary observation: If a parity check is adjacent to exactly one erased bit, then the value of that bit is the parity of the remaining bits in the check.
The decoder recovers this bit, removes it from the erasure and repeats.
To illustrate, we use a small example in Figure~\ref{fig:peeling-channel}.

Let $L \subseteq [n]$ be the set of erasure locations.
For a check $c \in C$, and $w \in \{0,1,...,\dC\}$, we say $c$ is a $w$-check if its \emph{weight}, defined as the number of neighbors in the erasure $\wt{\nbhd(c) \cap L}$, is $w$.

The $\peeling$ decoder is described in Algorithm~\ref{alg:peeling}. Linear time complexity can be achieved by maintaining a list of checks with a unique neighbor in the erasure and updating this list after each peeling step.
\begin{algorithm}[h]
    \begin{algorithmic}[0]
        \State \textbf{Input:} Syndrome $\synd(E)$ of an error $E$ and an erasure $L \subseteq [n]$ such that $E \subseteq L$.
    \end{algorithmic}
    \begin{algorithmic}[1]
        \State $\widehat{E} \gets \emptyset$, $\synd \gets \synd(E)$.
        \State Compute the weight of all checks and let $U \gets$ checks with weight $1$.
        \While{there exists $c \in U$}
            \State Let $v$ be the unique element of $\nbhd(c) \cap L$.
            \If{$\synd(c) = 1$}
                \State $\widehat{E} \gets \widehat{E} \cup \{v\}$ and flip $\synd(c')$ for all $c' \in \nbhd(v)$. \Comment{Recover $v$}
            \EndIf
            \State $L \gets L \setminus \{v\}$; update the weights of the checks in $\nbhd(v)$ and update $U$. \Comment{Peel $v$}
        \EndWhile
        \State \Return $\widehat{E}$ if $L = \emptyset$, and $\fail$ otherwise.
    \end{algorithmic}
    \caption{$\peeling$}
    \label{alg:peeling}
\end{algorithm}

Note that the size of the erasure $L$ decreases in each iteration.
Of course, $\peeling$ can fail prematurely if the set $U$ of unique neighbors is empty but the error is not yet corrected.
This is where we must demand constraints on the structure of the graph $G$ that defines the code $\ccode$.
In their seminal work, Sipser \& Spielman \cite{sipser1996expander} use an $(\alpha,\eps)$-expander graph $G$ to define a code $\ccode$.
For the moment, this means that the graph has the following property: For erasures $\env$ whose size is less than $\alpha n$, then each element of $\env$ has $(1-2\epsilon)\dV$ unique neighbors on average.
Thus if $\eps < 1/2$, then every non-empty erasure of at most $\alpha n$ bits has a unique neighbor.
Hence, we are guaranteed to be able to reduce the number of errors by at least 1 in every iteration of peeling.

\subsection{Quantum Expander Codes \& Small-Set-Peeling}
\label{subsec:sspeeling-motivation-overview}
We now move on to the quantum case.
Consider a $\dsl N,K,D\dsr$ quantum code $\qcode$ defined by a graph $\cG$ which specifies a set of parity checks $\pcs$ and generators $\gens$.
The parity checks $\pcs$ and generators $\gens$ serve to define two classical linear codes $\ccode_X$ and $\ccode_Z$ respectively: $\pcs$ gives the parity-checks for $\ccode_X$ and $\gens$ gives the parity-checks for $\ccode_Z$.
For $\qcode$ to be a valid quantum code, we require $\ccode_X^{\perp} \subseteq \ccode_Z$.

The set of errors in the quantum setting are partitioned into equivalence classes: Two errors are said to be equivalent if they differ by an element of $\ccode_Z^{\perp}$.
This changes the nature of the decoding problem for quantum codes relative to the classical setting: We only need to correct errors up to this equivalence, i.e.\ up to an element of $\ccode_Z^{\perp}$.

The problem of erasure correction is then framed as follows: Consider an erasure $\env \subseteq \qubits$.
We wish to design a decoder such that, for all errors $\err \subseteq \env$, the decoder accepts as input the syndrome $\synd(\err)$ and the envelope $\env$, and finds a correction $\corr$ such that $\err$ and $\corr$ only differ by an element of $\ccode_Z^{\perp}$.

\paragraph{Quantum Expander Codes.}
We define quantum expander codes formally in Section~\ref{subsec:background-quantum-expander}, but for now we give an informal definition.

\input{src_tikz/quadrants}

Let $G$ be a biregular, bipartite graph over $V \sqcup C$.
The quantum expander code $\qcode$ will be built out of the Cartesian product of two copies $G_1$, $G_2$ of $G$ \footnote{In principle, $G_1$ and $G_2$ can be different, but we keep our presentation simple.}.
The copies $G_1$ and $G_2$ are defined over $V_1 \sqcup C_1$ and $V_2 \sqcup C_2$ respectively.
The set of qubits is labeled $\qubits$ and can be partitioned as $\qubits = \qubits_V \sqcup \qubits_C$, where
\begin{align}
    \qubits_V := V_1 \times V_2~, \qquad \qubits_C := C_1 \times C_2~.
\end{align}
The $X$-type parity checks $\pcs$ and $Z$-type generators $\gens$ are associated with $V_1\times C_2$ and $C_1 \times V_2$ respectively.

This graph product is informally illustrated in Figure~\ref{fig:quadrants}.
The connectivity of $\cG$ is inherited from the connectivity of $G_1$ and $G_2$.
For example, if we consider the subgraph induced by fixing the first component $\nu_1 \in V_1$, then the graph induced by vertices $\{\nu_1\} \times (V_2 \sqcup C_2)$ is a copy of $G_2$.
In a similar way, fixing the second component yields a copy of $G_1$.

This graph product defines the sets $\pcs$ and $\gens$, and hence the quantum hypergraph product code $\qcode$.
 
\paragraph{Expansion (or its lack thereof).}
Even if $G$ is itself a good expander, the graph product $\cG$ is \emph{not} a very good expander.
In particular, there are very small sets with no unique neighbors.
To see such an example, recall that the set of generators $\gens$ is identified with $C_1 \times V_2$. 
Consider a generator $(c_1,v_2) \in \gens$ and its support 
\begin{align}
    \supp(c_1,v_2) = \Gamma(c_1) \times \{v_2\} \sqcup \{c_1\} \times \Gamma(v_2)~.
\end{align}
This set has \emph{no} unique neighbors.
We refer the reader to Figure~\ref{fig:supp-erasure}, where we present the local view of $(c_1,v_2)$, which includes the qubits $\supp(c_1,v_2)$ as well as the parity checks $(\nu_1,\zeta_2)$ that share support with $(c_1,v_2)$.
Every such parity check $(\nu_1,\zeta_2)$ satisfies $\nu_1 \in V_1$ and $\zeta_2 \in C_2$ and overlaps the support of $(c_1,v_2)$ at exactly 2 points.
This ensures that $\ccode_X^{\perp} \subseteq \ccode_Z$ and therefore represents a fundamental feature of quantum expander codes.
The failure of the unique-neighborhood property therefore explains the failure of the $\peeling$ decoder: It resolves one qubit at a time, and stalls on the supports of generators.

\input{src_tikz/supp-erasure}

We build up to $\sspeeling$ using two vignettes to intuit how we overcome the failure of the unique-neighborhood property.

\paragraph{Vignette \#1.} First, we consider a very simple scenario, that of an isolated erasure.
It builds on the example of the small set above in Figure~\ref{fig:supp-erasure} which shows that the unique-neighbor property fails for quantum expander codes.
As above, let $(c_1,v_2)$ specify a generator and consider an erasure $\env$ that is \emph{only} supported on $\supp(c_1,v_2)$ and nowhere else.
Let $\err$ be some error that is supported within $\env$.
The pair $(\err,\env)$ is isolated in that, to correct $\err$, we do not need to worry about the interaction of parity checks with errors outside $\supp(c_1,v_2)$.

We define pair checks to be the checks incident to the erasure on exactly two locations.
A pair check $(\nu_1,\zeta_2)$ in the local view of the generator $(c_1,v_2)$ meets $\supp(c_1,v_2)$ at the qubit $\qubit_V \in \nbhd(c_1)\times\{v_2\}$ and the qubit $\qubit_C \in \{c_1\}\times\nbhd(v_2)$
By themselves, pair checks cannot identify the error: For any value of the syndrome $\synd(\nu_1,\zeta_2)$ for a pair check $(\nu_1,\zeta_2)$, there are two possible errors in their support that are compatible:
\begin{align}
    \qubit_V \gets 0,\, \qubit_C \gets \synd(\nu_1,\zeta_2)~,\quad \text{or} \quad
    \qubit_V \gets 1,\, \qubit_C \gets 1 \oplus \synd(\nu_1,\zeta_2)~, 
\end{align}
where $\oplus$ represents addition modulo 2.
Once we choose the values for any pair $\qubit_V,\qubit_C$ adjacent to a pair check, then the remaining qubits are then fixed.
For example, consider the pair check $(\nu_1,\zeta_2')$ adjacent to $\qubit_C$.
If $\qubit_C$ is assigned a value (either it is decided to be the correction $\corr$ or not), then the $\qubit_V'$ which is the qubit in $\nbhd(c_1) \times \nbhd(\zeta_2')$ adjacent to the pair check $(\nu_1,\zeta_2')$ must have the correct assignment to satisfy the syndrome $\synd(\nu_1,\zeta_2')$.
This determines $\err$ up to complementing all of $\supp(c_1,v_2)$: The correction $\corr$ is either $\err$ or $\err \oplus \supp(c_1,v_2)$.

To see why having two solutions is not a fundamental obstacle, we refer to how the quantum decoding problem differs from its classical version: As discussed above, a correction $\corr$ need not be identical to the error $\err$; it is valid as long as $\err$ and $\corr$ are \emph{equivalent} up to an element of $\ccode_Z^{\perp}$.
In this case, $\supp(c_1,v_2)$ belongs to the basis of $\ccode_Z^{\perp}$ and both options for $\corr$ are valid corrections.

This motivates why $\sspeeling$ resolves subsets of the support of a single generator at once.

\paragraph{Vignette \#2.} Consider a general erasure $\env$ and suppose we apply $\peeling$ until we have removed all the unique neighbors to the erasure $\env$.
In what remains, if the only types of pair checks that appeared did so inside a generator whose \emph{entire} support belonged to the erasure, the decoding problem is easy: We would just repeat the argument in Vignette \#1.
Unfortunately, a general erasure $\env$ may only \emph{partially} overlap $\supp(c_1,v_2)$, so the complement $\err \oplus \supp(c_1,v_2)$ may contain qubits outside $\env$.
Nevertheless, considering the entire support $\supp(c_1,v_2)$ is an equally valid target for the decoder, as it is equivalent to $\err$ up to an element of $\ccode_Z^{\perp}$.
Motivated by this observation, we \emph{add} $\supp(c_1,v_2) \setminus \env$ to the erasure, and the support of the generator lies entirely inside the erasure.
We refer to Figure~\ref{fig:enlarge-erasure} to illustrate Vignette \#2.

\begin{figure}[ht]
    \centering
    \begin{tikzpicture}[every node/.style={font=\small}]
        \def\s{0.6}
        \node at (-2.25,3.5) {($a$)};
        \fill[pgray] (0,0) rectangle (5*\s,6*\s);
        \fill[ppeach] (0,0) rectangle (3*\s,4*\s);
        \draw[gridline] (0,0) grid[step=\s] (5*\s,6*\s);
        \draw[pline] (0,0) rectangle (5*\s,6*\s);
        \draw[tpeach, thick] (0,0) rectangle (3*\s,4*\s);
        \foreach \i in {1,...,4} \node[var, erased] at (-0.55,{(\i-0.5)*\s}) {};
        \foreach \i in {5,6} \node[var, known] at (-0.55,{(\i-0.5)*\s}) {};
        \foreach \j in {1,2,3} \node[var, erased] at ({(\j-0.5)*\s},-0.55) {};
        \foreach \j in {4,5} \node[var, known] at ({(\j-0.5)*\s},-0.55) {};
        \node[gennode] at (-0.55,-0.55) {};
        \node[tlav, anchor=north east,rotate=-45] at (-0.2,-1.2) {$(c_1,v_2)$};
        \draw[pbrace, trose] (-0.9,0) -- (-0.9,4*\s) node[midway, left=5pt, trose] {$\sub_V$};
        \draw[pbrace, trose] (3*\s,-0.9) -- (0,-0.9) node[midway, below=5pt, trose] {$\sub_C$};
        \draw[pbrace] (-0.9,4*\s) -- (-0.9,6*\s) node[midway, left=5pt, align=right] {not\\erased};
        \draw[pbrace] (5*\s,-0.9) -- (3*\s,-0.9) node[midway, below=5pt, align=center] {not\\erased};
        \begin{scope}[xshift=200]
        \node at (-2.25,3.5) {($b$)};
            \fill[ppeach] (0,0) rectangle (5*\s,6*\s);
            \draw[gridline] (0,0) grid[step=\s] (5*\s,6*\s);
            \draw[pline] (0,0) rectangle (5*\s,6*\s);
            
            \fill[white] (0,0) rectangle (3*\s,4*\s);
            \draw[gray] (0,0) grid[step=\s] (3*\s,4*\s);
            \draw[tpeach, thick] (3*\s,0) -- (3*\s,4*\s) -- (0,4*\s) -- (0,6*\s)--(5*\s,6*\s)--(5*\s,0)--(3*\s,0);
            \foreach \i in {1,...,4} \node[var, known] at (-0.55,{(\i-0.5)*\s}) {};
            \foreach \i in {5,6} \node[var, erased] at (-0.55,{(\i-0.5)*\s}) {};
            \foreach \j in {1,2,3} \node[var, known] at ({(\j-0.5)*\s},-0.55) {};
            \foreach \j in {4,5} \node[var, erased] at ({(\j-0.5)*\s},-0.55) {};
            \node[gennode] at (-0.55,-0.55) {};
            \node[tlav, anchor=north east,rotate=-45] at (-0.2,-1.2) {$(c_1,v_2)$};
            \draw[pbrace] (-0.9,0) -- (-0.9,4*\s) node[midway, left=5pt, align=right] {not\\ erased};
            \draw[pbrace] (3*\s,-0.9) -- (0,-0.9) node[midway, below=5pt,align=center] {not\\ erased};
        \end{scope}
    \end{tikzpicture}
    \caption{
        Vignette \#2 in the local view of $(c_1,v_2)$.
        ($a$) The erasure only partially overlaps $\supp(c_1,v_2)$.
        The checks between $\sub_V$ and $\sub_C$ (orange) have weight $(1,1)$.
        The gray checks are incident to at least two erasures.
        ($b$) By adding all of $\supp(c_1,v_2)$ to $\env$ and resolving $\sub$ using its pair checks, we obtain a new, smaller erasure.
        The white checks now correspond to satisfied checks.
    }
    \label{fig:enlarge-erasure}
\end{figure}

At first glance, this appears like we are going in the wrong direction: We have increased the size of the erasure.
However, we can now use the observation from Vignette \#1 to address this problem.
Let $\sub$ denote the subset of qubits adjacent to pair checks.
We fix the value of one of the qubits in $\sub$ arbitrarily.
Each pair check in the local view of $(c_1,v_2)$ now determines its second qubit.
This one choice then cascades and we resolve all qubits adjacent to pair checks at the cost of potentially introducing new errors into the code in the locations $\supp(\gen) \setminus \sub$.
We therefore must include $\supp(\gen)$ in the erasure.
However, we can subtract qubits in $\sub$ as they are now resolved.
In other words, we have exchanged $\sub$ for $\supp(\gen) \setminus \sub$.

\textbf{Why is this exchange valid?} When we pick a value for a qubit in $\sub$, we are changing the erasure $\env$ and the error $\err$ itself.
Let $\err'$ and $\env'$ be the error and the envelope after this process.
The new error $\err'$ may differ from the old one up to the support $\supp(c_1,v_2)$ of a stabilizer generator.
Therefore, any solution that we obtain for the new problem is equally valid.

\textbf{When is this exchange worth it?}
In this step, the erasure gains $\wt{\supp(c_1,v_2)\setminus\env}$ qubits and loses $\wt{\sub}$ qubits, so it shrinks as long as $\wt{\sub} > \wt{\supp(c_1,v_2)\setminus\env}$.
Our main technical lemma (Lemma~\ref{lem:prime-existence}) shows that expansion guarantees the existence of such a generator whenever $\peeling$ stalls.

\begin{algorithm}[h]
    \begin{algorithmic}[0]
        \State \textbf{Input:} Syndrome $\synd(\err)$ and erasure $\env \subseteq \qubits$.
    \end{algorithmic}
    \begin{algorithmic}[1]
        \While{$\env$ is not empty}
            \State Run classical $\peeling$ until it is no longer possible to do so.
            In each iteration of $\peeling$, $\corr$ grows by one qubit, the qubit $\qubit$ that is peeled.
            \State If no unique neighbors of the erasure exist, then find a pair $(\gen,\sub)$ such that $\sub \subseteq \supp(\gen) \cap \env$ and is more than half the support of $\gen$.
            \State Update $\env$ to include $\supp(\gen)$: $\env \gets \env \union \supp(\gen) \setminus \sub$.
            \State Find a subset $\add \subseteq \sub$ such that the pair checks adjacent to $\sub$ are satisfied.
            Add $\add$ to $\corr$.
        \EndWhile
        \State \Return $\corr$.
    \end{algorithmic}
    \caption{$\sspeeling$-$\mathtt{informal}$}
    \label{alg:sspeeling-informal}
\end{algorithm}

\subsection{Comparison with other erasure decoders}

For surface codes, Delfosse and Z\'emor~\cite{delfosse2020linear} gave a linear-time maximum-likelihood erasure decoder, which forms the second stage of the $\unionfind$ decoder~\cite{delfosse2021almost}.
It relies on the fact that every qubit of the surface code is adjacent to at most two checks of each type, so that qubits can be viewed as the edges of a graph whose vertices are the checks.
The decoder computes a spanning forest of the erased edges and peels it from its leaves: The check at a leaf is adjacent to only one erased edge of the forest, so its syndrome bit determines the value of that edge.
Each erased edge outside the forest completes a cycle of erased edges, and such a cycle is either a product of generators supported inside the erasure, or a logical operator, in which case no decoder can do better; either way, these edges can be set to $0$.
In other words, the only obstruction to peeling a surface code is a stabilizer that already lies inside the erasure.
This is no longer the case for quantum expander codes, whose qubits are adjacent to $\dV$ or $\dC$ checks of each type.

Several erasure decoders for quantum LDPC codes also use the fact that a generator supported on the erasure allows one qubit to be fixed arbitrarily.
Connolly, Londe, Leverrier, Delfosse consider the Pruned Peeling~\cite{connolly2022fast} decoder that fixes a qubit whenever the support of a single generator lies entirely inside the erasure.
Pech, G\"okduman, Yao, Pfister consider Dual Peeling~\cite{pech2026stabilizer} which performs row operations on the generator matrix to find products of generators that lie entirely inside the erasure.
These generators then complement the constraints coming from the parity checks.
In both cases, the stabilizer used must be supported on the erasure, and the erasure itself never grows from one iteration to the next.
On the other hand, $\sspeeling$ uses a single generator $\gen$ whose support may stick out of the erasure.

%% file: src_tikz/quadrants.tex
\begin{figure}[ht]
    \centering
    \begin{tikzpicture}[scale=0.85, every node/.style={font=\small}]
        \fill[pmint!45]  (0,4.4)   rectangle (4,8.4);   %
        \fill[pblue!45]  (4.4,4.4) rectangle (8.4,8.4); %
        \fill[plav!45]   (0,0)     rectangle (4,4);     %
        \fill[pmint!45]  (4.4,0)   rectangle (8.4,4);   %
        \draw[quad] (0,4.4) rectangle (4,8.4);
        \draw[quad] (4.4,4.4) rectangle (8.4,8.4);
        \draw[quad] (0,0) rectangle (4,4);
        \draw[quad] (4.4,0) rectangle (8.4,4);
        \node[tgreen, anchor=north west] at (0.05,8.35) {$\qubits_V$};
        \node[tblue,  anchor=north east] at (8.35,8.35) {$\pcs$};
        \node[tlav,   anchor=south west] at (0.05,0.05) {$\gens$};
        \node[tgreen, anchor=south east] at (8.35,0.05) {$\qubits_C$};
        \node at (-0.45,6.4) {$V_1$};
        \node at (-0.45,2.0) {$C_1$};
        \node at (2.0,8.8) {$V_2$};
        \node at (6.4,8.8) {$C_2$};

        \def\xv{2.6}   %
        \def\yc{1.6}   %
        \def\xz{5.4}   %
        \def\yn{6.4}   %

        \draw[tlav!60, dashed, thick] (\xv,0.2) -- (\xv,8.2);
        \draw[tlav!60, dashed, thick] (0.2,\yc) -- (8.2,\yc);
        \draw[tblue!60, dashed, thick] (0.2,\yn) -- (8.2,\yn);
        \draw[tblue!60, dashed, thick] (\xz,0.2) -- (\xz,8.2);

        \foreach \y in {5.2,7.6} \node[var, fill=plav] at (\xv,\y) {};
        \foreach \x in {6.6,7.8} \node[var, fill=plav] at (\x,\yc) {};
        \foreach \x in {0.8,3.5} \node[var, fill=pblue] at (\x,\yn) {};
        \foreach \y in {0.8,2.9} \node[var, fill=pblue] at (\xz,\y) {};
        \node[var, fill=prose, minimum size=3.4mm] (qv) at (\xv,\yn) {};
        \node[var, fill=prose, minimum size=3.4mm] (qc) at (\xz,\yc) {};
        \node[trose, anchor=south east] at (\xv,\yn) {$(\nu_1,v_2)$};
        \node[trose, anchor=north west] at (\xz,\yc) {$(c_1,\zeta_2)$};

        \node[gennode, minimum size=3.4mm] (g) at (\xv,\yc) {};
        \node[tlav, anchor=north east] at (\xv,\yc) {$(c_1,v_2)$};
        \node[chk, fill=pbluedeep, minimum size=3.2mm] (x) at (\xz,\yn) {};
        \node[tblue, anchor=south west] at (\xz,\yn) {$(\nu_1,\zeta_2)$};
    \end{tikzpicture}
    \caption{
        Quadrant view of the hypergraph product code~\cite{tillich2013quantum}.
        Qubits are identified with $\qubits_V = V_1 \times V_2$ and $\qubits_C = C_1 \times C_2$, parity checks with $V_1 \times C_2$, and generators with $C_1 \times V_2$.
        Each row is a copy of $G_2$ and each column is a copy of $G_1$.
        The support of the generator $(c_1,v_2)$ (purple, dashed) consists of $\nbhd(c_1) \times \{v_2\}$ in its column and $\{c_1\} \times \nbhd(v_2)$ in its row.
        The support of the check $(\nu_1,\zeta_2)$ (blue, dashed) consists of $\{\nu_1\} \times \nbhd(\zeta_2)$ in its row and $\nbhd(\nu_1) \times \{\zeta_2\}$ in its column.
        When $\nu_1 \in \nbhd(c_1)$ and $\zeta_2 \in \nbhd(v_2)$, the two supports overlap on exactly two qubits, $(\nu_1,v_2)$ and $(c_1,\zeta_2)$ (pink).
    }
    \label{fig:quadrants}
\end{figure}

%% file: src_tikz/supp-erasure.tex
\begin{figure}[ht]
    \centering
    \begin{tikzpicture}[every node/.style={font=\small}]
        \def\s{0.6}
        \fill[ppeach] (0,0) rectangle (5*\s,6*\s);
        \draw[gridline] (0,0) grid[step=\s] (5*\s,6*\s);
        \draw[pline] (0,0) rectangle (5*\s,6*\s);
        \foreach \i in {1,...,6} \node[var, erased] (r\i) at (-0.55,{(\i-0.5)*\s}) {};
        \foreach \j in {1,...,5} \node[var, erased] (c\j) at ({(\j-0.5)*\s},-0.55) {};
        \node[gennode] at (-0.55,-0.55) {};
        \node[tlav, anchor=north east,rotate=-45] at (-0.2,-1.2) {$(c_1,v_2)$};
        \node[rotate=90] at (-1.15,3*\s) {$\nbhd(c_1) \times \{v_2\}$};
        \node at (2.5*\s,-1.15) {$\{c_1\} \times \nbhd(v_2)$};

        \node[chk, fill=pbluedeep] (x1) at (1.5*\s,1.5*\s) {};
        \draw[thick] (x1) to[out=180,in=0] (r2);
        \draw[thick] (x1) to[out=-90,in=90] (c2);
        \node at (1.5*\s-0.25,1.5*\s + 0.5) {$(\nu_1,\zeta_2)$};
        \node[chk, fill=pbluedeep] (x2) at (1.5*\s,4.5*\s) {};
        \node at (1.5*\s-0.25,4.5*\s + 0.5) {$(\nu_1,\zeta_2')$};
        \draw[thick] (x2) to[out=180,in=0] (r5);
        \draw[thick] (x2) to[out=-60,in=60] (c2);
    \end{tikzpicture}
    \caption{
        The local view of a generator $(c_1,v_2)$ when the erasure is exactly $\supp(c_1,v_2)$ (pink).
        Every parity check in the local view is adjacent to exactly two erased qubits, one in its row and one in its column; this is shown for two checks.
        Hence $\supp(c_1,v_2)$ has no unique neighbors, and $\peeling$ cannot make progress.
    }
    \label{fig:supp-erasure}
\end{figure}

%% file: 03-background.tex
\section{Background \& Definitions}
\label{sec:background}

Let $G$ be a $(\dV,\dC)$-biregular, bipartite graph over $V \union C$.
The quantum code $\qcode$ is constructed using the graph $\cG = G_1 \times G_2$, the Cartesian product of $G_1$ and $G_2$, copies of $G$.
The set of qubits, denoted $\qubits$, is a disjoint union $\qubits = \qubits_V \sqcup \qubits_C$ where $\qubits_V = V_1 \times V_2$ and $\qubits_C = C_1 \times C_2$.
With this association, we can write $N = \wt{\qubits}$.
The $X$-type parity checks $\pcs$ and $Z$-type generators $\gens$ are associated with $V_1 \times C_2$ and $C_1 \times V_2$ respectively.

The graph $\cG$ inherits connectivity from the input graphs: An element $(a, b)$ is connected to $\{a\} \times \nbhd(b) \sqcup \nbhd(a) \times \{b\}$.

\subsection{Expander graphs}
Let $V = [n]$ and $C = [m]$, and $G$ be a bipartite graph over vertex set $V \union C$ that is biregular with left node degree $\dV$ and right node degree $\dC$.

\begin{definition}[Graph neighborhood]
For $S_V \subseteq V$ and $S_C \subseteq C$, let $E(S_V,S_C)$ denote the set of edges between $S_V$ and $S_C$.

For $S_V \subseteq V$, we denote by $\nbhd(S_V)$ the neighborhood of $S_V$
\begin{align}
    \nbhd(S_V) = \{c \in C\;|~ |E(\{c\}, S_V)| \geq 1\}~.
\end{align}
For $S_V \subseteq V $, $\unbhd(S_V)$ denotes the unique neighborhood of $S$
\begin{align}
    \unbhd(S_V) = \{c \in C \;|~ |E(\{c\}, S_V)| = 1\}~.
\end{align}
The definition of neighborhood of $S_C\subseteq C$ is analogous.
\end{definition}
\begin{definition}[Vertex expander]\label{def:exp}
We say $G$ is an $(\alpha_V, \epsilon_V)$ left vertex-expander if for all $S_V\subseteq V$,
\begin{align}
    |S_V| \leq \alpha_V n \implies |\nbhd(S_V)| \geq (1-\epsilon_V)\dV|S_V|~.
\end{align}
The graph $G$ is an $(\alpha_C,\epsilon_C)$ right vertex-expander if
\begin{align}
    |S_C| \leq \alpha_C m \Longrightarrow |\nbhd(S_C)| \geq (1-\epsilon_C)\dC |S_C|~.
\end{align}
The graph $G$ is an $(\alpha_V, \epsilon_V,\alpha_C,\epsilon_C)$ bidirectional vertex expander if it is an $(\alpha_V, \epsilon_V)$ left vertex-expander and $(\alpha_C,\epsilon_C)$ right vertex-expander.
\end{definition}
The parameter $\epsilon_V$ is related to the number of collisions between outgoing edges of $S_V\subset V$, i.e.\ vertices in $C$ where multiple edges from $S_V$ are incident.

We state the following well-known fact; see \cite{sipser1996expander}.
\begin{fact}
\label{fact:unbhd}
    For non-empty $S_V\subset V$, if $|\nbhd(S_V)| \geq (1-\epsilon_V)\dV|S_V|$ then also $|\unbhd(S_V)| \geq (1-2\epsilon_V)\dV|S_V|$.
    Consequently, there exists a variable node $v \in S_V$ such that
    \begin{align*}
        \wt{\nbhd(v) \cap \unbhd(S_V)} \geq (1-2\eps_V)\dV~.
    \end{align*}
    A similar claim can be made about $S_C \subseteq C$ if $\wt{S_V} \leq \alpha_C m$.
\end{fact}

\subsection{Classical expander codes}
Expander codes were introduced by Sipser and Spielman \cite{sipser1996expander}.  They are built from a good vertex expander, and defined as follows. 
\begin{definition}[Expander code]\label{def:exp-code}
    Let $G$ be a $(\dV,\dC)$ bi-regular bipartite graph over $V\sqcup C$, which is an $(\alpha_V,\epsilon_V)$ left-vertex expander.
    Denote $|V|=n,|C|=m$.
    
    The expander code induced by the graph $G$ is the subspace $\ccode\subset\set{0,1}^n$ such that for every $\ccodeword\in \ccode$ and $c\in \ccode$,
    \[ \bigoplus_{v\in\nbhd(c)} w_v = 0~, \]
    where $\oplus$ is the {\rm XOR} function.
\end{definition}

\subsection{Quantum codes}\label{sec:quantum-codes}

Given a classical code $\ccode \in \bbF_2^{N}$, we let $\ccode^{\perp} = \{\u:\; \forall \v \in \ccode,\; \sum_i u_i v_i = 0 \pmod{2} \}$ denote its dual.
An $\dsl N, K, D\dsr$ CSS quantum code $\qcode$ on a set of $N$ qubits $\qubits$ is specified by two classical codes $\ccode_Z,\ccode_X \subseteq \bbF_2^N$ that obey the relation $\ccode_X^{\perp} \subseteq \ccode_Z$.
The codes $\ccode_X^{\perp}$ and $\ccode_Z^{\perp}$ form the parity checks for the quantum code.
We let $\pcs$ and $\gens$ be a fixed basis for $\ccode_X^{\perp}$ and $\ccode_Z^{\perp}$ respectively.

The number of encoded qubits $K$ is defined as
\begin{align}
    K = \dim(\ccode_X/\ccode_Z^{\perp}) = \dim(\ccode_X) + \dim(\ccode_Z) - N~.
\end{align}
The distances $D_Z$ and $D_X$ are defined as
\begin{align}
    D_Z = \min\{ |\mathbf{e}| : \mathbf{e} \in \ccode_X \setminus \ccode_Z^{\perp}\}~, \quad
    D_X = \min\{ |\mathbf{e}| : \mathbf{e} \in \ccode_Z \setminus \ccode_X^{\perp}\}~.
\end{align}
The distance of the code $\qcode$ is defined as $D := \min(D_X,D_Z)$.
Throughout the rest of this paper, unless otherwise stated, we use $D$ to refer to the \emph{design distance}, a lower bound on the actual distance as the actual distance is not directly computable.

We identify the set of qubits $\qubits$ with the set $\{1,...,N\}$ and vectors $\mathbf{e} \in \bbF_2^N$ with their support $\err \subseteq \qubits$.
Addition of vectors mod $2$ corresponds to the symmetric difference of sets, denoted $\oplus$: For $\err_1, \err_2 \subseteq \qubits$, we have
$\err_1 \oplus \err_2 = [\err_1 \setminus \err_2] \union  [\err_2 \setminus \err_1]$.

For $\gen \in \gens$ and $\pc \in \pcs$, $\supp(\gen) \subseteq \qubits$ and $\supp(\pc) \subseteq \qubits$ denote the supports of the corresponding basis element.

It turns out that $X$ and $Z$ errors can be corrected separately.
As the two cases mirror each other, we will only consider $Z$-type errors.
The $X$-type parity checks are used to correct them and we shall refer to these as {parity checks}.
In this case, we are interested in the syndromes for the $X$-type parity checks.
For an error $\err$, we let $\synd(\err)$ denote the $X$-type syndrome (unsatisfied $X$-type parity checks):
\begin{align}
    \synd(\err) = \set{\pc | |\supp(\chi) \cap \err| \neq 0 \pmod{2}}~.
\end{align}
The $Z$-type parity checks are not treated as constraints in this view; rather they define $\ccode_Z^{\perp}$, the set of words equivalent to the trivial codeword.  We refer to the set $\gens$ of $Z$-type parity checks as \emph{generators}.

An erasure is specified by a set $\env \subseteq \qubits$.
With respect to an erasure $\env$, for parity check $\pc \in \pcs$, and non-negative weights $w_V \in \{0,...,\dV\}$ and $w_C \in \{0,...,\dC\}$, we say $\pc$ is a $(w_V,w_C)$-check if its weight is $(w_V, w_C)$ where the weight is defined as the tuple $(\wt{\supp(\pc) \cap \env_V},\wt{\supp(\pc) \cap \env_C})$.

\subsection{Quantum Expander Codes}
\label{subsec:background-quantum-expander}

Let $G$ be a $(\dV, \dC)$-biregular bipartite graph over vertex set $V \sqcup C$.
Suppose it is an $(\alpha_V,\epsilon_V,\alpha_C,\epsilon_C)$ bidirectional vertex-expander, with
$|V| = n$ and $|C| = m$ such that $n > m$.

The $[n,k,d]$ code $\ccode(G)$ is the code defined in Definition~\ref{def:exp-code}, where $V$ and $C$ are associated with bits and parity checks respectively.
The $[m,\tk,\td]$ code $\tC(G)$ is the code associated with the graph $G$, in which $C$ and $V$ swap roles: $V$ is associated with parity checks and $C$ with the bits.
Suppose $\ccode(G)$ and $\tC(G)$ have distance $d$ and $\td$ respectively.
For convenience, we say that $d = \infty$ ($\td = \infty$) if $k = 0$ ($\tk = 0$).

The hypergraph product code $\qcode$ an $\dsl N,K,D\dsr$ code where $N = n^2 + m^2$, $K = (k)^2 + (\tk)^2$ and distance $D = \min(d,\td)$.
For the proof of these facts, we point the reader to the original paper by Tillich and Z\'emor \cite{tillich2013quantum}.
We proceed to describe the code construction.
Throughout this paper, we will assume $\dV > \dC$.

\begin{definition}
Let $G_1$ and $G_2$ denote two copies of a $(\dV,\dC)$-biregular, bipartite graph $G$ over the vertex set $V \sqcup C$.
The hypergraph product code $\qcode$ is constructed from the graph product $\cG := G_1 \times G_2$, where,
\begin{enumerate}
    \item The set of qubits, denoted $\qubits$, is associated with $\qubits_V \sqcup \qubits_C$, where $\qubits_V = V_1 \times V_2 $ and $\qubits_C = C_1 \times C_2$.
    \item The set of $Z$ stabilizer generators, denoted $\gens$, is associated with $C_1 \times V_2$.
    We shall let $\gen$ denote a generator and write $\gen \sim (c_1,v_2) \in C_1 \times V_2$ to say $\gen$ is identified with the tuple $(c_1,v_2)$.
    The support $\supp(\gen)$ of the generator $\gen$ is 
    \begin{equation}
        \supp(\gen) = \{(\nu_1,v_2): \nu_1 \in \nbhd(c_1)\} \union \{(c_1,\zeta_2): \zeta_2 \in \nbhd(v_2)\}~.
    \end{equation}
    \item The set of $X$ stabilizer generators, denoted $\pcs$, is associated with $V_1 \times C_2$.
    For $\nu_1 \in V_1$ and $\zeta_2 \in C_2$, the support $\supp(\pc)$ of a parity check $\pc \sim (\nu_1,\zeta_2)$ is 
    \begin{equation}
        \supp(\pc) = \{(\nu_1,v_2) : v_2 \in \nbhd(\zeta_2)\} \union \{(c_1,\zeta_2) : c \in \nbhd(\nu_1)\} ~.
    \end{equation}
\end{enumerate}
\end{definition}
We refer to Fig.\ \ref{fig:quadrants} for an illustration.

Throughout, we use Roman letters for $C_1$ and $V_2$, and Greek letters for $C_2$ and $V_1$.  Thus, generators are Roman tuples like $(c_1,v_2)$, while parity checks are Greek tuples like $(\nu_1, \zeta_2)$.

We note that in this case, the codes $\ccode_Z^{\perp}$ and $\ccode_X^{\perp}$ are given by
\begin{align*}
    \ccode_Z^{\perp} = {\rm span}(\supp(c_1,v_2) : c_1 \in C_1, v_2 \in V_2)~, \quad
    \ccode_X^{\perp} = {\rm span}(\supp(\nu_1,\zeta_2) : \nu_1 \in V_1, \zeta_2 \in C_2)~.
\end{align*}

If we ignore the generators $\gens$, we get a bipartite graph between the qubits $\qubits$ and then $X$-type parity checks $\pcs$ (that is, the induced graph of $\cG$ on the vertices $\qubits \sqcup \pcs$).  We denote this graph by $\cG_X$.

For a set of qubits $\qset\subseteq\qubits$, let $\qset_V = \qset\cap \qubits_V$ and $\qset_C = \qset\cap \qubits_C$.

For clarity, we use different symbols the neighborhood in the classical graph $G$ and in the quantum graph $\cG$.
In the classical setting, we use $\nbhd$, while in the quantum setting, we use $\qnbhd$.  Formally, we have the following definitions.
\begin{definition}[Quantum neighborhood]\label{def:qgraph}
Let $G=(V\cup C,E)$ be a graph and $\cG_{X}$ be the graph induced by qubits $\qubits$ and parity checks $\pcs$.
For a qubit $\qubit \in \qubits$, we denote its neighborhood by $\qnbhd(\qubit)$, defined as follows:
\begin{align*}
    &\qnbhd(\set{\nu_1,v_2}) = \set{(\nu_1,\zeta_2)| \zeta_2\in\nbhd(v_2)}~, \quad 
    &\qnbhd(\set{c_1, \zeta_2}) = \set{(\nu_1,\zeta_2)| \nu_1\in\nbhd(c_1)}~,
\end{align*}
where $\nbhd$ is the neighborhood in the graph $G$.
This naturally extends to a set of qubits: For $\qset \subseteq \qubits$,
\begin{align}
    \qnbhd(\qset) = \cup_{\qubit \in\qset}\qnbhd(\qubit)~.
\end{align}
\end{definition}

We shall be interested in two subsets of the neighborhood of a set $\qset$, the unique neighborhood and pair neighborhood respectively.

\begin{definition}[Unique and pair neighbors]
Let $\qset \subseteq \qubits$, and write $\qset_V = \qset \cap \qubits_V$ and $\qset_C = \qset \cap \qubits_C$.
The \emph{weight} of a check $\pc$ with respect to $\qset$ is the pair $(\wt{\supp(\pc) \cap \qset_V}, \wt{\supp(\pc) \cap \qset_C})$.
The \emph{unique neighbors} and \emph{pair neighbors} of $\qset$ are, respectively,
\begin{align}
    \uqnbhd(\qset) &= \set{\pc \in \pcs \,:\, \pc \text{ has weight } (1,0) \text{ or } (0,1)}~,\\
    \pqnbhd(\qset) &= \set{\pc \in \pcs \,:\, \pc \text{ has weight } (1,1)}~.
\end{align}
\end{definition}

\subsection{Projections of sets}

In this section, we introduce some notation for referring to projections of sets on to the component graphs $G_1$ and $G_2$.
This notation will be used for all projections throughout this paper.
We refer to Figure~\ref{fig:projections} for a schematic.

\input{src_tikz/projections}

\begin{definition}[Projection]\label{def:proj}
Let $\cG$ be the Cartesian product of $G_1$ and $G_2$, and consider a set of qubits \(\qset \subseteq \qubits\).

For all $\nu_1 \in V_1$ and all $v_2 \in V_2$, let $\qsetvtwo\subset V_1$, $\qsetvone\subset V_2$ be the projections on to $G_1$ and $G_2$ respectively.
\begin{align*}
    \qsetvtwo = \set{\nu_1' | (\nu_1',v_2) \in \qset_V}~, \qquad \qsetvone = \set{v_2' | (\nu_1,v_2') \in \qset_V}~.
\end{align*}
We obtain $\qsetVone\subset V_1$ and $\qsetVtwo\subset V_2$ as the union of sets
\begin{align}\label{eq:proj}
    \qsetVone = \bigcup_{v_2} \qsetvtwo~,\qquad \qsetVtwo = \bigcup_{\nu_1} \qsetvone~.
\end{align}
Similarly, for all $c_1 \in C_1$ and $\zeta_2 \in C_2$, let $\qsetcone\subset C_2$ and $\qsetctwo\subset C_1$ denote the projections on to $G_1$ and $G_2$ respectively:
\begin{align}
    \qsetcone = \set{\zeta_2' | (c_1,\zeta_2') \in \qset_C}~,\quad
    \qsetzetatwo = \set{c_1' | (c_1', \zeta_2) \in \qset_C}~.
\end{align}
We obtain $\qsetCone\subset C_1$ and $\qsetCtwo\subset C_2$ as the union of sets
\begin{align}    
    \qsetCtwo = \bigcup_{c_1 \in C_1} \qsetcone~,\quad\qsetCone = \bigcup_{\zeta_2 \in C_2} \qsetzetatwo~.
\end{align}
\end{definition}

\subsection{Local View}

In addition to the global view shown in Fig.~\ref{fig:projections}, we will also consider a local view which describes the connectivity of the graph from the point-of-view of a single generator $\gen \sim (c_1,v_2)$.
This is shown in Fig.~\ref{fig:local-view}.

\input{src_tikz/local-view}

\begin{obs}
\label{obs:local-view}
    Let $\gen \sim (c_1,v_2)$ be a generator.
    \begin{enumerate}
        \item Every check adjacent to a qubit of $\supp(\gen)$ is of the form $\pc \sim (\nu_1,\zeta_2)$ with $\nu_1 \in \nbhd(c_1)$ and $\zeta_2 \in \nbhd(v_2)$, and it meets $\supp(\gen)$ in exactly the two qubits $(\nu_1,v_2)$ and $(c_1,\zeta_2)$.
        \item Consequently, for $\sub \subseteq \supp(\gen)$, the pair neighborhood $\pqnbhd(\sub)$ consists of exactly one check for each pair $(\qubit_V,\qubit_C) \in \sub_V \times \sub_C$, and $\wt{\pqnbhd(\sub)} = \wt{\sub_V}\wt{\sub_C}$.
    \end{enumerate}
\end{obs}
\begin{proof}
    We prove each claim in turn.
    \begin{enumerate}
        \item A qubit $(\nu_1,v_2) \in \nbhd(c_1) \times \{v_2\}$ is adjacent to the checks $(\nu_1,\zeta_2)$ with $\zeta_2 \in \nbhd(v_2)$, and a qubit $(c_1,\zeta_2) \in \{c_1\} \times \nbhd(v_2)$ is adjacent to the checks $(\nu_1,\zeta_2)$ with $\nu_1 \in \nbhd(c_1)$.
        In both cases, the check is $(\nu_1,\zeta_2)$ with $\nu_1 \in \nbhd(c_1)$ and $\zeta_2 \in \nbhd(v_2)$.
        The support of such a check is $\{\nu_1\} \times \nbhd(\zeta_2) \sqcup \nbhd(\nu_1) \times \{\zeta_2\}$, which meets $\nbhd(c_1) \times \{v_2\}$ only in $(\nu_1,v_2)$, and $\{c_1\} \times \nbhd(v_2)$ only in $(c_1,\zeta_2)$.
        \item By Claim 1, a check has weight $(1,1)$ with respect to $\sub$ if and only if it is the check $(\nu_1,\zeta_2)$ with $(\nu_1,v_2) \in \sub_V$ and $(c_1,\zeta_2) \in \sub_C$.
        Distinct pairs give distinct checks, as the check determines $\nu_1$ and $\zeta_2$.
    \end{enumerate}
    This completes the proof.
\end{proof}

\subsection{Distance}

For completeness, we include a statement about the distance $D$.
As stated in Section~\ref{sec:quantum-codes}, we use $D$ to refer to a lower bound on the actual distance of the quantum expander code.
This distance is guaranteed by the expansion properties of the underlying graphs $G_1$ and $G_2$.

\begin{lemma}
\label{lem:distance}
    Let $G$ be a $(\dV,\dC)$-biregular, $(\alpha_V,\eps_V,\alpha_C,\eps_C)$ bidirectional expander with $\eps_V,\eps_C < 1/2$.
    Then $d > \alpha_V n$ and $\td > \alpha_C m$.
    Consequently, the hypergraph product code $\qcode$ has distance at least
    \begin{align}
        D := \min(\alpha_V n, \alpha_C m) + 1~.
    \end{align}
\end{lemma}
\begin{proof}
    Let $S \subseteq V$ be the support of a non-zero codeword of $\ccode(G)$, and suppose that $\wt{S} \leq \alpha_V n$.
    By Fact~\ref{fact:unbhd}, $\wt{\unbhd(S)} \geq (1-2\eps_V)\dV\wt{S} > 0$.
    A check in $\unbhd(S)$ is adjacent to exactly one bit of $S$, so it is violated, a contradiction.
    Hence, as $d$ is integer, $d > \alpha_V n$.
    The same argument applied to $\tC(G)$, using right expansion, gives $\td > \alpha_C m$.
    Finally, the distance of $\qcode$ is at least $\min(d,\td)$~\cite{tillich2013quantum}.
\end{proof}

%% file: src_tikz/projections.tex
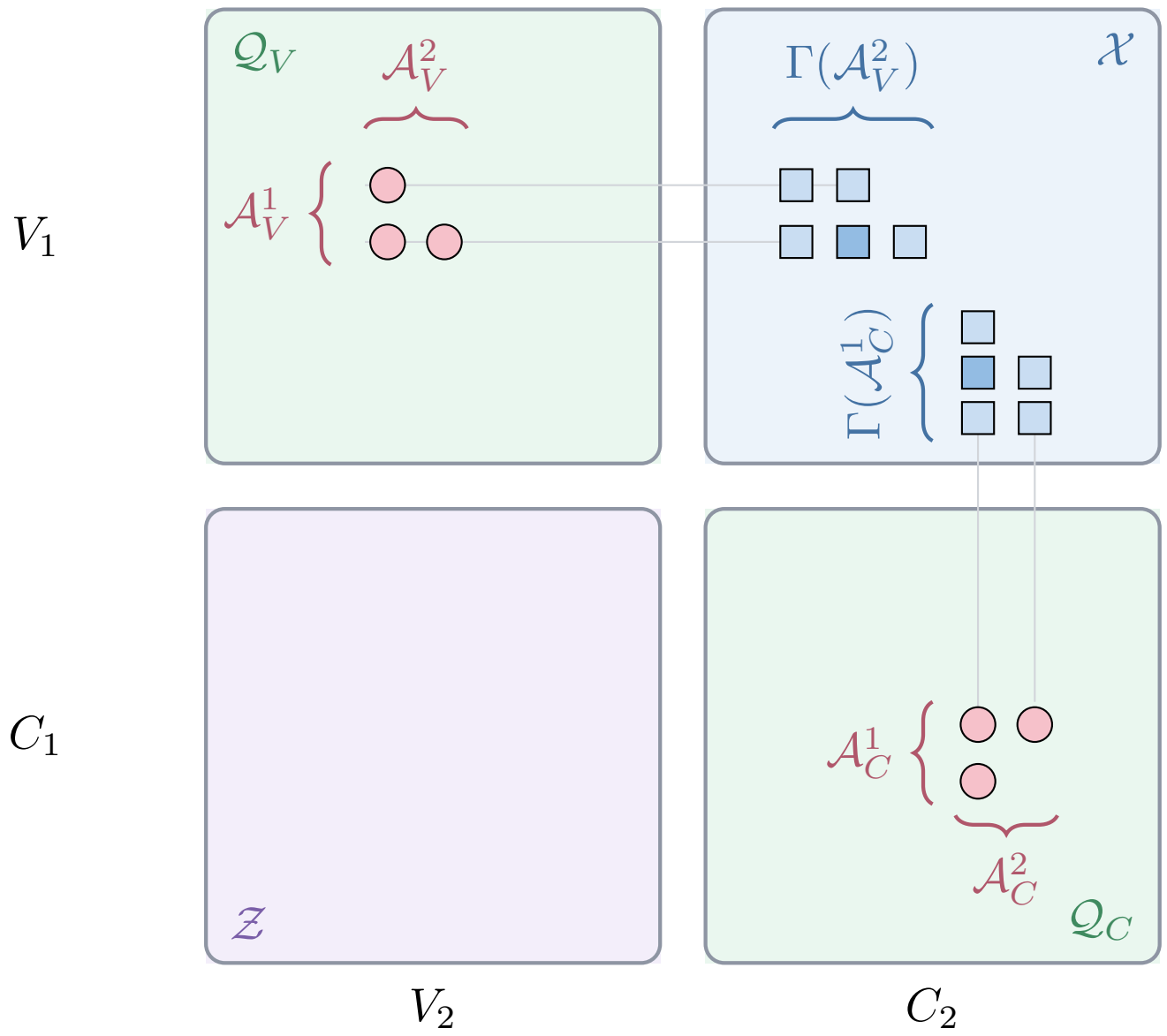
\begin{figure}[h]
\centering
\begin{tikzpicture}[scale=0.85, every node/.style={font=\small}]
    \fill[pmint!35]  (0,4.4)   rectangle (4,8.4);
    \fill[pblue!35]  (4.4,4.4) rectangle (8.4,8.4);
    \fill[plav!35]   (0,0)     rectangle (4,4);
    \fill[pmint!35]  (4.4,0)   rectangle (8.4,4);
    \draw[quad] (0,4.4) rectangle (4,8.4);
    \draw[quad] (4.4,4.4) rectangle (8.4,8.4);
    \draw[quad] (0,0) rectangle (4,4);
    \draw[quad] (4.4,0) rectangle (8.4,4);
    \node at (-1.5,6.4) {$V_1$};
    \node at (-1.5,2.0) {$C_1$};
    \node at (2.0,-0.4) {$V_2$};
    \node at (6.4,-0.4) {$C_2$};
    \node[tgreen, anchor=north west] at (0.05,8.35) {$\qubits_V$};
    \node[tblue,  anchor=north east] at (8.35,8.35) {$\pcs$};
    \node[tlav,   anchor=south west] at (0.05,0.05) {$\gens$};
    \node[tgreen, anchor=south east] at (8.35,0.05) {$\qubits_C$};

    \begin{scope}[yshift=10]
    \draw[pline!40] (1.4,6.0) -- (5.2,6.0);
    \draw[pline!40] (1.4,6.5) -- (5.7,6.5);
    \node[var, erased] at (1.6,6.5) {};
    \node[var, erased] at (1.6,6.0) {};
    \node[var, erased] at (2.1,6.0) {};
    \draw[pbrace, trose] (1.4,7.0) -- (2.3,7.0) node[midway, above=4pt, trose] {$\qsetVtwo$};
    \draw[pbrace, trose] (1.1,5.8) -- (1.1,6.7) node[midway, left=4pt, trose] {$\qsetVone$};
    \node[chk] at (5.2,6.5) {};
    \node[chk] at (5.7,6.5) {};
    \node[chk] at (5.2,6.0) {};
    \node[chk, unsat] at (5.7,6.0) {};
    \node[chk] at (6.2,6.0) {};
    \draw[pbrace, tblue] (5.0,7.0) -- (6.4,7.0) node[midway, above=4pt, tblue] {$\nbhd(\qsetVtwo)$};
    \end{scope}

    \draw[pline!40] (6.8,2.1) -- (6.8,4.8);
    \draw[pline!40] (7.3,2.3) -- (7.3,4.8);
    \node[var, erased] at (6.8,2.1) {};
    \node[var, erased] at (7.3,2.1) {};
    \node[var, erased] at (6.8,1.6) {};
    \draw[pbrace, trose] (7.5,1.3) -- (6.6,1.3) node[midway, below=4pt, trose] {$\qsetCtwo$};
    \draw[pbrace, trose] (6.4,1.4) -- (6.4,2.3) node[midway, left=4pt, trose] {$\qsetCone$};
    \node[chk] at (6.8,4.8) {};
    \node[chk] at (7.3,4.8) {};
    \node[chk, unsat] at (6.8,5.2) {};
    \node[chk] at (7.3,5.2) {};
    \node[chk] at (6.8,5.6) {};
    \draw[pbrace, tblue] (6.4,4.6) -- (6.4,5.8) node[midway, xshift=-14pt, tblue,rotate=90] {$\nbhd(\qsetCone)$};
\end{tikzpicture}
\caption{
    Projections of a set $\qset$ of qubits (pink) in the quadrant view of Fig.~\ref{fig:quadrants}.
    The set $\qset_V \subseteq V_1 \times V_2$ further projects onto $\qsetVone \subseteq V_1$ (its rows) and $\qsetVtwo \subseteq V_2$ (its columns); the set $\qset_C \subseteq C_1 \times C_2$ further projects onto $\qsetCone \subseteq C_1$ and $\qsetCtwo \subseteq C_2$.
    A qubit $(\nu_1,v_2)$ is adjacent to the checks $\{\nu_1\} \times \nbhd(v_2)$ in its row, so $\qnbhd(\qset_V)$ lies in the rows $\qsetVone$ and the columns $\nbhd(\qsetVtwo)$.
    A qubit $(c_1,\zeta_2)$ is adjacent to the checks $\nbhd(c_1) \times \{\zeta_2\}$ in its column, so $\qnbhd(\qset_C)$ lies in the columns $\qsetCtwo$ and the rows $\nbhd(\qsetCone)$.
    Darker checks are adjacent to two qubits of $\qset$.
}
\label{fig:projections}
\end{figure}

%% file: src_tikz/local-view.tex
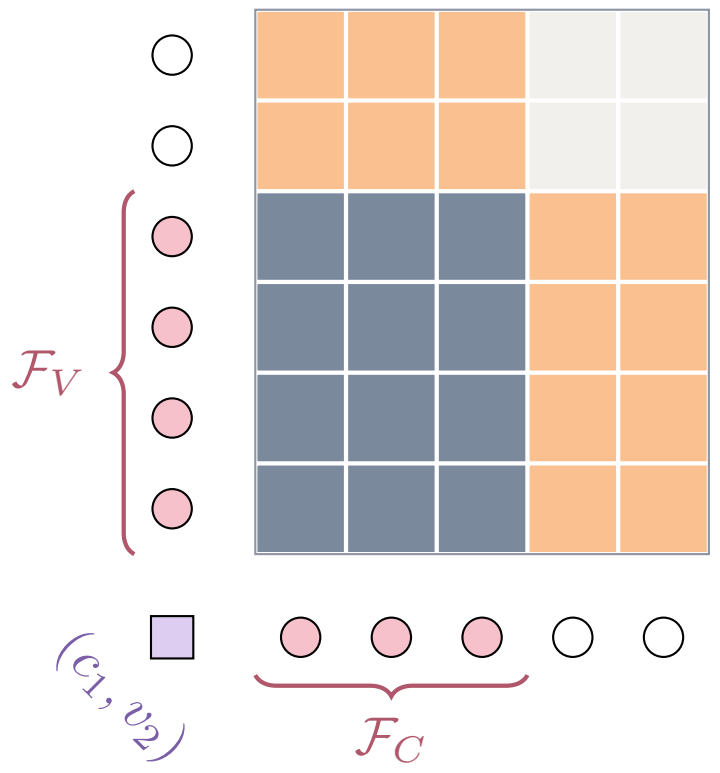
\begin{figure}[ht]
    \centering
    \begin{tikzpicture}[every node/.style={font=\small}]
        \def\s{0.6}
        \fill[ppeachdeep] (0,0) rectangle (5*\s,6*\s);
        \fill[tblue,opacity=0.7] (0,0) rectangle (3*\s,4*\s);
        \fill[pgray] (3*\s,4*\s) rectangle (5*\s,6*\s);
        \draw[gridline] (0,0) grid[step=\s] (5*\s,6*\s);
        \draw[pline] (0,0) rectangle (5*\s,6*\s);
        \foreach \i in {1,...,4} \node[var, erased] at (-0.55,{(\i-0.5)*\s}) {};
        \foreach \i in {5,6} \node[var, known] at (-0.55,{(\i-0.5)*\s}) {};
        \foreach \j in {1,2,3} \node[var, erased] at ({(\j-0.5)*\s},-0.55) {};
        \foreach \j in {4,5} \node[var, known] at ({(\j-0.5)*\s},-0.55) {};
        \node[gennode] at (-0.55,-0.55) {};
        \node[tlav, anchor=north east,rotate=-45] at (-0.2,-1.2) {$(c_1,v_2)$};
        \draw[pbrace, trose] (3*\s,-0.8) -- (0,-0.8) node[midway, below=4pt, trose] {$\add_C$};
        \draw[pbrace, trose] (-0.8,0*\s) -- (-0.8,4*\s) node[midway, left=5pt, trose, align=left] {$\add_V$};
        One highlighted check
    \end{tikzpicture}
    \caption{
        The local view of a generator $(c_1,v_2)$.
        Rows are indexed by the qubits $\nbhd(c_1) \times \{v_2\}$ and columns by the qubits $\{c_1\} \times \nbhd(v_2)$.
        Qubits in the set $\add \subseteq \supp(c_1,v_2)$ are pink.
        The cells in blue are the pair checks with respect to the set $\add$.
        Orange cells correspond to checks incident to an odd number of qubits of $\add$ and are the unique neighbors of $\add$.
        Gray cells correspond to checks not incident to $\add$.
    }
    \label{fig:local-view}
\end{figure}

%% file: 04-erasure_decoding.tex
\section[Small-Set-Peeling and its analysis]{$\sspeeling$ \& its analysis}
\label{sec:decoder}

In this section, we present the decoder $\sspeeling$ and prove Theorem~\ref{thm:main}.
Throughout, $\qcode$ is the hypergraph product of two copies $G_1$ and $G_2$ of a $(\dV,\dC)$-biregular, bipartite graph $G$ over $V \sqcup C$ such that it is a $(\alpha_V,\eps_V,\alpha_C,\eps_C)$-bidirectional expander $G$, as described in Section~\ref{subsec:background-quantum-expander}.
The decoder is given an erasure $\env \subseteq \qubits$ and the syndrome $\synd(\err)$ of an unknown error $\err \subseteq \env$.
For an erasure $\env$, we write $\unq := \uqnbhd(\env)$ for its unique neighbors, the checks of weight $(1,0)$ or $(0,1)$, and $\pair := \pqnbhd(\env)$ for its pair neighbors, the checks of weight $(1,1)$.

$\sspeeling$ repeatedly shrinks the erasure while keeping track of an error that is equivalent to $\err$ and supported inside it.
Each iteration makes one of two moves.
If the erasure has a unique neighbor $\pc \in \unq$, then, as in classical $\peeling$, the syndrome bit of $\pc$ reveals the value of its only erased qubit, and we remove this qubit from the erasure.
Otherwise, we look for a \emph{valid pair} $(\gen,\sub)$ (Definition~\ref{def:valid-pair}): A generator $\gen$, together with a set $\sub$ of erased qubits in its support whose pair neighbors all have weight $(1,1)$, and which is larger than the part of $\supp(\gen)$ outside the erasure.
These checks determine the error on $\sub$ up to complementing all of $\sub$, i.e.\ up to adding $\supp(\gen)$.
We resolve this ambiguity arbitrarily, remove $\sub$ from the erasure, and add $\supp(\gen) \setminus \env$ to it.

The rest of this section is organized as follows.
Section~\ref{subsec:decoder} defines valid pairs and presents the decoder.
Section~\ref{subsec:existence} proves our main technical lemma: Expansion guarantees that one of the two moves is always available (Lemma~\ref{lem:prime-existence}).
Section~\ref{subsec:single-iteration} shows that each move shrinks the erasure (Lemma~\ref{lem:env-shrinks}) while keeping the error correctable (Lemma~\ref{lem:err-remains-correctable}).
Section~\ref{subsec:proof-main} combines these to prove Theorem~\ref{thm:main}, including the running time.
We begin with a formal version of Theorem~\ref{thm:maininformal} below.

\begin{restatable}{theorem}{main}
\label{thm:main}
    Consider a bipartite, $(\dV,\dC)$-biregular graph $G$ on vertex set $V \sqcup C$, where $V = [n]$, $C = [m]$ such that $G$ is a $(\alpha_V,\alpha_C,\epsilon_V,\epsilon_C)$-bidirectional expander with $\eps_V, \eps_C < 1/4$.
    Furthermore, we assume $G$ is unbalanced, $n > m$ or equivalently that $\dC > \dV$.
    Let $\qcode$ be a $\dsl N,K, D\dsr$ quantum expander code obtained via the hypergraph product of two copies of $G$.
    There exists a decoder $\sspeeling$ such that, for all $\env \subseteq \qubits$ such that $\wt{\env} \leq \min(\alpha_V n, \alpha_C m)$, we have the following guarantee.
    For any ($Z$-type) error $\err \subseteq \env$, $\sspeeling$ takes as input the syndrome $\synd(\err)$ and the envelope $\env$ as input, and returns a correction $\corr$ in $O_{\dV,\dC}(\wt{\env})$ time such that $\err \oplus \corr \in \ccode_Z^{\perp}$.
\end{restatable}

\subsection{Valid Pairs}
\label{subsec:valid-pair}
We begin by defining the main ingredient that allows us to overcome the obstacles faced by the classical $\peeling$ decoder.
We define a generator $\gen$ and a large subset $\sub \subseteq \supp(\gen) \cap \env$ such that $\sub$ is adjacent to many pair checks within the local view of $\gen$.
Informally, such generators straddle the envelope, i.e.\ their support sits partially within the envelope.
Such a generator $\gen$ and subset $\sub$ will then allow us to find a correction when $\peeling$ exhausts the set of unique neighbors of $\env$.
When we find such a pair $(\gen,\sub)$, we replace $\env$ with $\left(\env \union \supp(\gen)\right) \setminus \sub$, i.e. we swap $\sub$ within $\env$ for $\supp(\gen) \setminus \sub$.

\begin{definition}
\label{def:valid-pair}
With respect to the erasure $\env$, a valid pair $(\gen, \sub)$ consists of a generator $\gen \in \gens$ and a subset $\sub \subseteq \env\, \cap\, \supp(\gen)$ such that all of the following conditions hold:
\begin{enumerate}
    \item \(\sub\) has non-trivial support on both sets of qubits, $\sub_V \neq \emptyset$,  $\sub_C \neq \emptyset$.
    \item The pair neighborhood of \(\sub\) is contained in the pair neighborhood of the erasure $\pqnbhd(\sub) \subseteq \pair$.
    \item \(\sub\) is large, $\wt{\sub} > \wt{\supp(\gen) \setminus \env}$.
\end{enumerate}
\end{definition}

\input{src_tikz/valid-pair}

By Observation~\ref{obs:local-view}, condition 2 of Definition~\ref{def:valid-pair} says that for all pairs $(\nu_1,v_2) \in \sub_V$ and $(c_1,\zeta_2) \in \sub_C$, the qubits $(\nu_1,v_2)$ and $(c_1,\zeta_2)$ are the \emph{only} erased qubits in the support of $\pc \sim (\nu_1,\zeta_2)$.

Condition 3 is what makes the erasure shrink (Lemma~\ref{lem:env-shrinks}).
As $\sub \subseteq \env$, we have $\supp(\gen) \setminus \env \subseteq \supp(\gen) \setminus \sub$, and $\wt{\supp(\gen)} = \dV + \dC$, so condition 3 holds whenever $\wt{\sub} > (\dV + \dC)/2$.
The valid pairs we construct in Lemma~\ref{lem:prime-existence} satisfy this stronger condition.

\subsection{Decoder}
\label{subsec:decoder}

Let $\env \subseteq \qubits$ be an erasure and $\err \subseteq \env$ be an error in its support.
The decoder $\sspeeling$ is given the syndrome $\synd(\err)$ and the erasure $\env$ and tasked with finding a correction $\corr \subseteq \qubits$ such that $\err \oplus \corr \in \ccode_Z^{\perp}$.
The decoder is an iterative decoding algorithm.
At its heart is the subroutine $\iter$, presented in Algorithm~\ref{alg:iter}, which is called every iteration.

Consider the $i$\textsuperscript{th} iteration of $\sspeeling$.
The subroutine $\iter$ takes as input an erasure $\env_{i}$ and syndrome $\synd_{i}$ and returns three objects, $\add_{i}$, $\sub_{i}$ and $\gen_i$.

For the $(i+1)$\textsuperscript{th} iteration, it then uses as input
\begin{align}
    \synd_{i+1} = \synd_{i} \oplus \synd(\add_{i})~,\quad \env_{i+1} = (\env_{i} \union \supp(\gen_i)) \setminus \sub_i~.
\end{align}

\begin{algorithm}[h]
    \begin{algorithmic}[0]
        \State \textbf{Input:} $\synd = \synd(\err), \env$
    \end{algorithmic}
    \begin{algorithmic}[1]
        \State $\synd_0 \gets \synd$, $\env_0 \gets \env$, $I \gets \wt{\env}$, $\corr_0 \gets \emptyset$.
        \State Compute weight of all checks and push them into $\pcs(w_V, w_C)$, an unordered set of checks with weight $(w_V, w_C)$.
        \For{$0 \leq i < I $}
        \State $\gen_i, \,\sub_i, \,\add_i \gets \iter(\synd_{i}, \env_{i})$
        \State $\corr_{i+1} \gets \corr_{i} \oplus \add_i$,\, $\synd_{i+1} \gets \synd_i \oplus \synd(\add_i)$,\, $\env_{i+1} \gets \left(\env_{i} \union \supp(\gen_i)\right) \setminus \sub_{i}$
        \State Update $\pcs(w_V,w_C)$ by recomputing the weights of parity checks adjacent to $\sub_i \union (\supp(\gen_i) \setminus \env_i)$.
        \State Update syndrome of parity checks adjacent to $\add_i$.
        \EndFor
        \If{$\env_{I} = \emptyset$}
            \State \Return $\corr_{I}$
        \Else
            \State \Return $\fail$
        \EndIf
    \end{algorithmic}
    \caption{$\sspeeling$}
    \label{alg:decode}
\end{algorithm}

\begin{algorithm}[h]
    \begin{algorithmic}[0]\onehalfspacing
        \State \textbf{Input:} Syndrome $\synd_{i}$ and erasure $\env_{i}$.
    \end{algorithmic}
    \begin{algorithmic}[1]\onehalfspacing
        \State $\unq \gets \uqnbhd(\env_i)$, $\pair \gets \pqnbhd(\env_i)$.
        \State $\gen_i \gets 0$, $\sub_i, \add_i \gets \emptyset$.\Comment{$\supp(0) = \emptyset$}
        \If{there exists a $\pc \in \unq$}
                \State $\sub_{i} \gets \supp(\pc) \cap \env_i$.
                \State $\add_{i} \gets \supp(\pc) \cap \env_i$ if and only if $\synd_i(\pc) = 1$.
        \ElsIf{there exists valid pair $(\gen, \sub)$ with respect to $\env_i$}
            \State $\gen_i \gets \gen$, $\sub_{i} \gets \sub$.
            \State $\add_i \gets \Fill(\synd_i, \gen_i, \sub_i)$.
        \EndIf
        \State \Return $\gen_i, \sub_{i}, \add_{i}$
    \end{algorithmic}
    \caption{$\iter$}
    \label{alg:iter}
\end{algorithm}

Algorithm~\ref{alg:iter} is stated without data structures, and its test ``there exists a valid pair'' should not be read as a search over all generators.
We return to this in Lemma~\ref{lem:runtime} where we account for the runtime.

\input{src_tikz/fill}

The subroutine $\Fill$ (Algorithm~\ref{alg:Fill}) determines the error on $\sub$ up to complementing all of $\sub$.
By Observation~\ref{obs:local-view}, the checks in $\pqnbhd(\sub)$ are in bijection with the pairs in $\sub_V \times \sub_C$.
As each of these checks has weight $(1,1)$, its syndrome bit is the parity of the error on the two endpoints of its edge.
Once the value at one qubit is chosen, the remaining qubits in $\sub$ automatically fall into $\corr$ or its complement.
$\Fill$ uses a fixed pair of qubits $\qubit_0 \in \sub_V$ and $\qubit_1 \in \sub_C$ (Fig.~\ref{fig:fill}) to come up with an assignment.
The value at $\qubit_0$ is set to $0$; this is the arbitrary choice that is paid for by the generator $\gen$ in Lemma~\ref{lem:err-remains-correctable}.

\begin{algorithm}[h]
    \begin{algorithmic}[0]\onehalfspacing
        \State \textbf{Input:} $\synd$, and a valid pair $(\gen,\sub)$
    \end{algorithmic}
    \begin{algorithmic}[1]\onehalfspacing
        \State Order qubits lexicographically; let $\qubit_0$ be the first qubit of $\sub_V$ and $\qubit_1$ be the first qubit of $\sub_C$.
        \State Let $\pc_0$ be the unique parity check corresponding to $\qubit_0$ and $\qubit_1$.
        \State $\add \gets \emptyset$.
        \For{$\qubit_C \in \sub_C$}
            \State Let $\pc$ be the unique parity check corresponding to $\qubit_0$ and $\qubit_C$.
            \If{$\synd(\pc) = 1$}
                \State $\add \gets \add \union \{\qubit_C\}$.
            \EndIf
        \EndFor
        \For{$\qubit_V \in \sub_V \setminus \{\qubit_0\}$}
            \State Let $\pc$ be the unique parity check corresponding to $\qubit_1$ and $\qubit_V$.
            \If{$\synd(\pc) \oplus \synd(\pc_0) = 1$}
                \State $\add \gets \add \union \{\qubit_V\}$.
            \EndIf
        \EndFor
        \State \Return $\add$
    \end{algorithmic}
    \caption{$\Fill$}
    \label{alg:Fill}
\end{algorithm}

\subsection{Existence \& structure of valid pairs}
\label{subsec:existence}

\begin{lemma}
\label{lem:prime-existence}
    Suppose $\eps_V, \eps_C < 1/4$ and $\env \neq \emptyset$.
    If $\wt{\envV} \leq \alpha_V n$ and $\wt{\envC} \leq \alpha_C m$, then either there exists a unique neighbor $\unq \neq \emptyset$ or there exists a valid pair $(\gen,\sub)$.
\end{lemma}
\begin{proof}
Let $\envV = \env \cap \qubits_V$ and $\envC = \env \cap \qubits_C$.
Consider the projection $\envVtwo$ of $\envV$:
\begin{align}
    \envVtwo := \{ v_2 \,:\, \text{There exists } \nu_1 \in V_1 \text{ such that } (\nu_1,v_2) \in \envV \}~.
\end{align}

\begin{claim}
    If $\envVtwo = \emptyset$, then there exists a check $\pc \in \unq$.
\end{claim}
\begin{claimproof}
    If $\envVtwo = \emptyset$, then $\envV = \emptyset$.
    Consider the projection $\envCone$ of $\envC$:
    \begin{align}
        \envCone := \{ c_1 \,:\, \text{There exists } \zeta_2 \in C_2 \text{ such that } (c_1,\zeta_2) \in \envC \}~.
    \end{align}
    Per the assumption, $\wt{\envCone} \leq \wt{\envC} \leq \alpha_C m$.
    Thus, by Fact~\ref{fact:unbhd}, there will exist at least one $c_1 \in \envCone$ such that
    \begin{align}
        \wt{\nbhd(c_1) \cap \unbhd(\envCone)} \geq (1-2\eps_C)\dC~.
    \end{align}
    Let $\zeta_2 \in C_2$ such that $(c_1,\zeta_2) \in \envC$.
    For all $\nu_1 \in \nbhd(c_1)\cap \unbhd(\envCone)$, the check $(\nu_1,\zeta_2)$ belongs to $\unq$ and is incident to $\envC$ only at $(c_1,\zeta_2)$.
\end{claimproof}

Going forward, suppose $\envVtwo$ is non-empty and, per the assumption, satisfies $\wt{\envVtwo} \leq \wt{\envV} \leq \wt{\env} \leq \alpha_V n$.
Therefore, there exists $v_2 \in \envVtwo$ such that
\begin{align}
    T_2 := \nbhd(v_2) \cap \unbhd(\envVtwo) \subseteq C_2~, \qquad \wt{T_2} \geq (1-2\eps_V)\dV~.
\end{align}

Let $\indC \subseteq \envC$ be the subset of $\envC$ induced by $\unbhd(\envVtwo) \cap \nbhd(v_2)$:
\begin{align}
    \indC = \envC \cap \left[C_1 \times \left( \unbhd(\envVtwo) \cap \nbhd(v_2)\right) \right]~.
\end{align}

Let $\indCone$ be the projection of $\indC$ on to its first component.
\begin{align}
    \indCone := \{ c_1 \,:\, \text{There exists } \zeta_2 \in C_2 \text{ such that } (c_1,\zeta_2) \in \indC \}~.
\end{align}

\begin{claim}
    If $\indCone = \emptyset$, then there exists a parity check $\pc \in \unq$.
\end{claim}
\begin{claimproof}
    Suppose $\indCone = \emptyset$.
    As $v_2 \in \envVtwo$, there must exist at least one $\nu_1 \in V_1$ such that $(\nu_1,v_2) \in \envV$.
    Any element $\zeta_2 \in \unbhd(\envVtwo) \cap \nbhd(v_2)$ defines a check $(\nu_1,\zeta_2)$ in $\unq$.
    This is because it is incident to the erasure only at $(\nu_1,v_2) \in \qubits_V$.
\end{claimproof}

Going forward, we suppose that $\indCone$ is non-empty.
It satisfies $\wt{\indCone} \leq \wt{\envC} \leq \alpha_C m$.
Therefore, by Fact~\ref{fact:unbhd}, there exists $c_1 \in \indCone$ such that
\begin{align}
    T_1 := \nbhd(c_1) \cap \unbhd(\indCone) \subseteq V_1~,\qquad \wt{T_1} \geq (1-2\eps_C)\dC
\end{align}
Consider the generator $\gen \sim (c_1,v_2)$ and the subset of its support
\begin{align}
    \sub := \left(T_1 \times \{v_2\}\right) \sqcup \left(\{c_1\} \times T_2\right)~,
\end{align}
with $\sub_V = T_1 \times \{v_2\}$ and $\sub_C = \{c_1\} \times T_2$.
By Observation~\ref{obs:local-view}, $\pqnbhd(\sub)$ consists of the checks $(\nu_1,\zeta_2)$ with $\nu_1 \in T_1$ and $\zeta_2 \in T_2$.

\begin{claim}
\label{claim:local-weights}
    Let $\nu_1 \in T_1$, $\zeta_2 \in T_2$ and $\pc \sim (\nu_1,\zeta_2)$.
    Then $\supp(\pc) \cap \env \subseteq \{(\nu_1,v_2), (c_1,\zeta_2)\}$.
\end{claim}
\begin{claimproof}
    The qubits of $\supp(\pc)$ in $\qubits_V$ are $(\nu_1,v_2')$ with $v_2' \in \nbhd(\zeta_2)$.
    If such a qubit is erased, then $v_2' \in \envVtwo$.
    As $\zeta_2 \in \unbhd(\envVtwo)$ is adjacent to $v_2 \in \envVtwo$, this forces $v_2' = v_2$.
    The qubits of $\supp(\pc)$ in $\qubits_C$ are $(c_1',\zeta_2)$ with $c_1' \in \nbhd(\nu_1)$.
    If such a qubit is erased, then it belongs to $\indC$ as $\zeta_2 \in T_2$, so $c_1' \in \indCone$.
    As $\nu_1 \in \unbhd(\indCone)$ is adjacent to $c_1 \in \indCone$, this forces $c_1' = c_1$.
\end{claimproof}

In words, the checks in $\pqnbhd(\sub)$ interact with the erasure \emph{only} inside $\supp(\gen)$.
Next, we consider the $\star$-conditions:
\begin{enumerate}
    \item $\{c_1\} \times T_2 \subseteq \env$.
    \item $T_1 \times \{v_2\} \subseteq \env$.
\end{enumerate}
It is illustrated in Fig.~\ref{fig:star-condition}.

\input{src_tikz/star-condition}

\begin{claim}
    Suppose $\unq = \emptyset$.
    Then both $\star$-conditions hold.
\end{claim}
\begin{claimproof}
    By Claim~\ref{claim:local-weights}, every check $(\nu_1,\zeta_2)$ with $\nu_1 \in T_1$ and $\zeta_2 \in T_2$ has weight $(w_V,w_C)$ with $w_V = \indicator{(\nu_1,v_2) \in \env}$ and $w_C = \indicator{(c_1,\zeta_2) \in \env}$.
    As $\unq = \emptyset$, no such check has weight $(1,0)$ or $(0,1)$, so
    \begin{align}
    \label{eq:star-equivalence}
        (\nu_1,v_2) \in \env \iff (c_1,\zeta_2) \in \env \qquad \text{for all } \nu_1 \in T_1,\ \zeta_2 \in T_2~.
    \end{align}
    As $c_1 \in \indCone$, there exists $\zeta_2^\star$ with $(c_1,\zeta_2^\star) \in \indC \subseteq \env$, and $\zeta_2^\star \in T_2$ by the definition of $\indC$.
    Applying Eq.~\eqref{eq:star-equivalence} with $\zeta_2 = \zeta_2^\star$ gives $(\nu_1,v_2) \in \env$ for all $\nu_1 \in T_1$, which is the second $\star$-condition.
    As $\wt{T_1} \geq (1-2\eps_C)\dC > 0$, we may fix some $\nu_1^\star \in T_1$.
    Applying Eq.~\eqref{eq:star-equivalence} with $\nu_1 = \nu_1^\star$ gives $(c_1,\zeta_2) \in \env$ for all $\zeta_2 \in T_2$, which is the first $\star$-condition.
\end{claimproof}

Note that the erased qubit witnessing $v_2 \in \envVtwo$ need not lie in $\supp(\gen)$; the erasures in $\sub_V$ are instead forced by the assumption $\unq = \emptyset$.

We conclude the proof.
Suppose $\unq = \emptyset$.
By the claim above, $\sub \subseteq \env \cap \supp(\gen)$.
Both $\sub_V$ and $\sub_C$ are non-empty, and by Claim~\ref{claim:local-weights}, every check in $\pqnbhd(\sub)$ has weight $(1,1)$.
Finally, as $\eps_V,\eps_C < 1/4$,
\begin{align}
    \wt{\sub} = \wt{T_1} + \wt{T_2} \geq (1-2\eps_C)\dC + (1-2\eps_V)\dV > \frac{\dV+\dC}{2}~,
\end{align}
which implies condition 3 of Definition~\ref{def:valid-pair}.
Hence $(\gen,\sub)$ is a valid pair.
\end{proof}

\subsection{Analyzing a single iteration}
\label{subsec:single-iteration}

First, we consider how the erasure $\env_{i}$ itself changes from one iteration to the next.

\begin{lemma}[Erasure shrinks]
\label{lem:env-shrinks}
    Consider an error $\err_{i}$ and an erasure $\env_{i}$ such that $\err_{i} \subseteq \env_{i}$.
    Let $\gen_i, \sub_i, \add_i \gets \iter(\synd_i, \env_i)$ and suppose $\sub_i \neq \emptyset$.
    Then $\env_{i+1} = \env_{i} \union \supp(\gen_i) \setminus \sub_i$ obeys $\wt{\env_{i+1}} \leq \wt{\env_{i}} - 1$.
\end{lemma}
\begin{proof}
     We consider the size of the erasure $\env_{i}$ in both cases.
    
     \emph{Case 1.} There exists a unique neighbor $\pc \in \unq$ of the erasure.
     
     In this case, $\env_{i+1} \gets \env_{i} \setminus \sub_{i}$.
     By definition, all checks in $\unq$ overlap the erasure in one and only one location, i.e.\ $\sub_{i} = \supp(\pc) \cap \env$ is a singleton set.
     Therefore,
     \begin{align}
         \wt{\env_{i+1}} &= \wt{\env_{i} \setminus \sub_{i}}\\
         &\leq \wt{\env_{i}} - 1~.
     \end{align}

     \emph{Case 2.} There are no unique neighbors of the erasure $\unq = \emptyset$ and there exists a valid pair $(\gen_i, \sub_i)$.

     By Definition~\ref{def:valid-pair}, the subset $\sub_i \subseteq \supp(\gen)$ obeys $\wt{\sub_i \cap \supp(\gen)} > \wt{\supp(\gen) \setminus \env_i}$.
     Therefore,
     \begin{align}
         \wt{\env_{i+1}} &= \wt{\env_{i} \union \supp(\gen) \setminus \sub_i}\\
         &\leq \wt{\env_{i}} - \wt{\sub_{i} \cap \supp(\gen)} + \wt{\supp(\gen_i) \setminus \env_{i}}\\
         &< \wt{\env_{i}}~.
     \end{align}

     This completes the proof.
\end{proof}

Next, we analyze how the error associated with this erasure evolves.
While the following lemma may look dense at first glance, it merely says that our choice for $\add$ in the $i$\textsuperscript{th} iteration is either the error $\err$ within $\sub$ or the equivalent $\err \oplus \supp(\gen)$ within $\sub$.

\begin{lemma}
\label{lem:err-remains-correctable}
    Consider an error $\err_{i}$ and an erasure $\env_{i}$.
    Let $\synd_i$ be the syndrome input to $\iter$.
    Suppose $\sub_i, \add_i \gets \iter(\synd_i, \env_i)$ and suppose $\sub_i \neq \emptyset$.
    If $\err_{i} \subseteq \env_{i}$ and $\synd_i = \synd(\err_{i})$, then there exists $\delta_i \in \{0,1\}$ such that for new error \(\err_{i+1}\)
    \begin{align}
        \err_{i+1} := \err_{i} \oplus \add_i \oplus \delta_i \supp(\gen_i)~,
    \end{align}
    the error \(\err_{i+1}\) satisfies:
    \begin{enumerate}
        \item The updated envelope contains the new error $\err_{i+1} \subseteq \env_{i+1}$.
        \item The updated syndrome corresponds to the new error $\synd(\err_{i}) \oplus \synd(\add_{i}) = \synd(\err_{i+1})$.
        \item The new error is equivalent to the previous error after correction $\err_{i} \oplus \add_{i} \oplus \err_{i+1} \in \ccode_{Z}^{\perp}$.
    \end{enumerate}
\end{lemma}
\begin{proof}
    The heart of the proof is the following claim, which says that the correction $\add_i$ agrees with the error $\err_i$ on $\sub_i$, up to complementing all of $\sub_i$.

    \begin{claim}
    There exists $\delta_i \in \{0,1\}$ such that $\err_{i} \cap \sub_i = \add_{i} \oplus \delta_i \sub_i$.
    Furthermore, $\delta_i = 0$ if $\gen_i = 0$.
    \end{claim}
    \begin{claimproof}
    \emph{Case 1.} There exists a unique neighbor $\pc \in \unq$.
    
    Then $\sub_i = \{\qubit\}$, where $\qubit$ is the only erased qubit in $\supp(\pc)$.
    As $\err_i \subseteq \env_i$, we have $\synd_i(\pc) = \indicator{\qubit \in \err_i}$.
    By construction, $\add_i = \{\qubit\}$ if and only if $\synd_i(\pc) = 1$, so $\err_i \cap \sub_i = \add_i$.

    \emph{Case 2.} No unique neighbor exists $\unq = \emptyset$ and there exists a valid pair $(\gen,\sub)$.

    Consider a pair $\qubit_V' \in \sub_V$ and $\qubit_C' \in \sub_C$ and let $\pc'$ be the unique parity check adjacent to both qubits.
    By condition 2 of Definition~\ref{def:valid-pair} and Observation~\ref{obs:local-view}, the qubits $\qubit_V'$ and $\qubit_C'$ are the only erased qubits in $\supp(\pc')$.
    As $\err_i \subseteq \env_i$, the syndrome of \(\pc'\) is
    \begin{align}
    \label{eq:pair-relation}
        \synd_i(\pc') = \indicator{\qubit_V' \in \err_i} \oplus \indicator{\qubit_C' \in \err_i}~.
    \end{align}
    Now let $\qubit_0$ and $\qubit_1$ be the qubits chosen in Algorithm~\ref{alg:Fill}, let $\pc_0$ be the parity check incident to both of them, and let $\delta_i = \indicator{\qubit_0 \in \err_i}$.
    Using Eq.~\eqref{eq:pair-relation} we now analyze \(\add_i\) qubit-by-qubit as constructed by Algorithm~\ref{alg:Fill} and show its relation to \(\err_i\):
    \begin{enumerate}
        \item For $\qubit_0$: $\indicator{\qubit_0 \in \err_i} = \delta_i = \delta_i \oplus \indicator{\qubit_0 \in \add_i}$, as $\qubit_0 \notin \add_i$.
        \item Consider $\qubit_0$ and some $\qubit_C \in \sub_C$ with unique incident parity check $\pc$.
        \begin{align}
            \indicator{\qubit_C \in \err_i} = \indicator{\qubit_0 \in \err_i} \oplus \synd_i(\pc) = \delta_i \oplus \indicator{\qubit_C \in \add_i}~.
        \end{align}
        \item Consider some $\qubit_V \in \sub_V \setminus \{\qubit_0\}$ and $\qubit_1$  with unique incident parity check $\pc$.
        \begin{align}
            \indicator{\qubit_V \in \err_i} = \indicator{\qubit_1 \in \err_i} \oplus \synd_i(\pc) = \delta_i \oplus \synd_i(\pc_0) \oplus \synd_i(\pc) = \delta_i \oplus \indicator{\qubit_V \in \add_i}~.
        \end{align}
    \end{enumerate}
    As the correction satisfies $\add_i \subseteq \sub_i$, this shows the error must satisfy $\err_i \cap \sub_i = \add_i \oplus \delta_i \sub_i$.
    \end{claimproof}

    Note that the decoder does not know $\delta_i$; it is only used in the analysis.
    We now verify the three items using \(\delta_i\) from the claim.
    \begin{enumerate}
        \item First consider $\sub_i \cap \err_{i+1}$: In case 2, $\sub_i \subseteq \supp(\gen_i)$, and in case 1, $\delta_i = 0$.
        In both cases, using the claim, \(\err_{i+1}\) cannot intersect \(\sub_i\)
        \begin{align}
            \err_{i+1} \cap \sub_i = (\err_i \cap \sub_i) \oplus \add_i \oplus \delta_i \sub_i = \emptyset~.
        \end{align}
        Since \(\err_{i+1}\) must be contained in the old erasure and generator $\err_{i+1} \subseteq \err_i \union \add_i \union \supp(\gen_i) \subseteq \env_i \union \supp(\gen_i)$, and having established that \(\err_{i+1} \cap \sub_i = \emptyset\), the item follows from $\err_{i+1} \subseteq \left(\env_i \union \supp(\gen_i)\right) \setminus \sub_i = \env_{i+1}$.
        \item By linearity of the syndrome, $\synd(\err_{i+1}) = \synd(\err_i) \oplus \synd(\add_i) \oplus \delta_i \synd(\supp(\gen_i)) = \synd_i \oplus \synd(\add_i)$.
        \item $\err_{i} \oplus \add_{i} \oplus \err_{i+1} = \delta_i \supp(\gen_i)$, which is either $\emptyset$ or the support of a generator, and in both cases belongs to $\ccode_{Z}^{\perp}$.
    \end{enumerate}
    This completes the proof.
\end{proof}

\subsection{Proof of Thm.~\ref{thm:main}}
\label{subsec:proof-main}

\subsubsection{Correctness}
First, we show that the algorithm corrects the error in $O(\wt{\env})$ iterations.

Consider $\env \subseteq \qubits$ such that $\wt{\env} \leq \min(\alpha_V n, \alpha_C m)$ and an error $\err \subseteq \env$.
Let $\err_0 = \err$, and consider the iterations of Algorithm~\ref{alg:decode}.
We show by induction on $i$ that, as long as $\env_i \neq \emptyset$, the following hold:
\begin{enumerate*}[label=(\roman*)]
    \item $\wt{\env_i} \leq \wt{\env} - i$,
    \item $\err_i \subseteq \env_i$ and $\synd_i = \synd(\err_i)$, and
    \item $\err \oplus \corr_i \oplus \err_i \in \ccode_Z^{\perp}$.
\end{enumerate*}
These hold for $i = 0$ as $\corr_0 = \emptyset$.

Suppose they hold for some $i$ with $\env_i \neq \emptyset$.
By (i), $\wt{\env_{i,V}} \leq \wt{\env_i} \leq \alpha_V n$ and $\wt{\env_{i,C}} \leq \wt{\env_i} \leq \alpha_C m$.
By Lemma~\ref{lem:prime-existence}, either $\unq \neq \emptyset$ or there exists a valid pair, so $\iter$ returns $\sub_i \neq \emptyset$ and the decoder does not output $\fail$.
Lemma~\ref{lem:env-shrinks} then gives (i) for $i+1$, and Lemma~\ref{lem:err-remains-correctable} gives (ii) for $i+1$ together with $\err_i \oplus \add_i \oplus \err_{i+1} \in \ccode_Z^{\perp}$.
Adding this to (iii) for $i$, and using $\corr_{i+1} = \corr_i \oplus \add_i$, gives (iii) for $i+1$.

By (i), the loop terminates after some $I \leq \wt{\env}$ iterations with $\env_I = \emptyset$.
By (ii), $\err_I = \emptyset$, and by (iii), the output $\corr = \corr_I$ satisfies $\err \oplus \corr \in \ccode_Z^{\perp}$.

\subsubsection{Time complexity}
We assume that the erasure $\env$ is given as a list and the syndrome $\synd(\err)$ as the list of its unsatisfied checks.
As $\err \subseteq \env$, every unsatisfied check is adjacent to $\env$, so this list has size at most $\dC\wt{\env}$.

\begin{lemma}
\label{lem:runtime}
    $\sspeeling$ can be implemented in $O_{\dV,\dC}(\wt{\env})$ time.
\end{lemma}
\begin{proof}
    The implementation maintains the following data, where $\env$ refers to the current erasure:
    \begin{enumerate}
        \item For every check adjacent to $\env$, its weight $(w_V,w_C)$ and its current syndrome bit.
        \item A queue $\unq$ containing the checks of weight $(1,0)$ or $(0,1)$.
        \item A queue $\gens_{{\rm possible}}$ of generators to be examined, containing every generator whose local view has changed since it was last examined.
    \end{enumerate}
    \emph{Initialization.}
    Every check adjacent to $\env$ is one of the at most $\dC\wt{\env}$ checks adjacent to some qubit of $\env$, and its weight and syndrome bit are computed in $O(\dV+\dC)$ time.
    $\gens_{{\rm possible}}$ is initialized with the generators whose support meets $\env$, of which there are at most $\dC\wt{\env}$.

    \emph{Finding a move.}
    If $\unq$ is non-empty, we pop a check and, if it still has weight $(1,0)$ or $(0,1)$, perform a peeling step.
    Otherwise, we pop generators from $\gens_{{\rm possible}}$ until one of them is part of a valid pair.
    For a popped generator $\gen$, deciding whether there exists $\sub \subseteq \supp(\gen) \cap \env$ such that $(\gen,\sub)$ is a valid pair takes $O(2^{\dV+\dC}\dV\dC)$ time by examining every subset of $\supp(\gen)$.
    A generator that fails this test is discarded.
    
    \emph{Total.}
    Initialization pushes at most $\max(\dV,\dC)\wt{\env}$ generators.
    An iteration changes the erasure on at most $\dV + \dC$ qubits, hence the weights of at most $(\dV+\dC)\max(\dV,\dC)$ checks, and pushes at most $(\dV+\dC)\max(\dV,\dC)\dV\dC$ generators.
    By Lemma~\ref{lem:env-shrinks}, there are at most $\wt{\env}$ iterations, so the total number of pushes is $O_{\dV,\dC}(\wt{\env})$.
    Each pop removes an element that was pushed earlier, so the total number of pops, summed over all iterations, is also $O_{\dV,\dC}(\wt{\env})$, even though a single search may take longer.
    The same argument applies to $\unq$.
    Finally, if $\env \neq \emptyset$ and $\unq$ has no check of weight $(1,0)$ or $(0,1)$, then by Lemma~\ref{lem:prime-existence} some generator is part of a valid pair, and by the invariant it is in $\gens_{{\rm possible}}$.
    Hence a search never empties the queue without finding a move, and the total running time is $O_{\dV,\dC}(\wt{\env})$.
\end{proof}

\subsection{Proof of Corollary~\ref{cor:error-decoder}}

The work of \cite{krishna2024viderman} proved the following result:

\begin{theorem}
    Let $\mathcal{H}$ be an $\dsl N,K,D \dsr$ hypergraph product code constructed from a $(\dV,\dC)$-biregular bipartite graph $G$ that is a $(\alpha_V,\epsilon_V,\alpha_C,\epsilon_C)$ bidirectional expander such that $\epsilon_V, \epsilon_C < 1/10$.

    Let $\err$ be a (reduced) error pattern with weight at most 
    \[ |\err| < \gamma \cdot D - \Delta_V, \]
    where 
    $ \gamma = \frac{\Delta_V}{\Delta_C} \cdot \frac{1 - 10 \epsilon}{4}$, and let $\synd(\err)$ be the syndrome arising from $\err$.
    There exists an algorithm $\ssfind$ such that, given $\synd(\err)$, runs in time $O_{\Delta_V, \Delta_C}(|\err|)$, and returns an envelope $\env \subseteq \mathcal{Q}$, so that $\err \subseteq \env$, and 
    \[ |\env| \leq \frac{1}{\gamma} \cdot |\err|. \]
\end{theorem}

Given $\synd(\err)$, $\ssfind$ returns an envelope $\env \supseteq \err$ in time $O_{\dV,\dC}(\wt{\err})$, and
\begin{align}
    \wt{\env} \leq \frac{\wt{\err}}{\gamma} < D - \frac{\dV}{\gamma} < D~.
\end{align}
As $\eps_V, \eps_C < 1/10 < 1/4$, Theorem~\ref{thm:main} applies to the pair $(\synd(\err),\env)$: $\sspeeling$ returns $\corr$ with $\err \oplus \corr \in \ccode_Z^{\perp}$ in time $O_{\dV,\dC}(\wt{\env}) = O_{\dV,\dC}(\wt{\err}/\gamma)$.
As $\gamma$ depends only on $\dV$, $\dC$ and $\eps$, the total running time is $O_{\dV,\dC,\eps}(\wt{\err})$.

%% file: src_tikz/valid-pair.tex
\begin{figure}[ht]
    \centering
    \begin{tikzpicture}[every node/.style={font=\small}]
        \def\s{0.6}
        \fill[pgray] (0,0) rectangle (5*\s,6*\s);
        \fill[ppeach] (1*\s,1*\s) rectangle (4*\s,5*\s);
        \draw[gridline] (0,0) grid[step=\s] (5*\s,6*\s);
        \draw[pline] (0,0) rectangle (5*\s,6*\s);
        \draw[tpeach, thick] (1*\s,1*\s) rectangle (4*\s,5*\s);
        \draw[tpeach, dashed] (-0.25,1*\s) -- (5*\s,1*\s) (-0.25,5*\s) -- (5*\s,5*\s);
        \draw[tpeach, dashed] (1*\s,-0.25) -- (1*\s,6*\s) (4*\s,-0.25) -- (4*\s,6*\s);
        \node[var, known] at (-0.55,0.5*\s) {};
        \foreach \i in {2,...,5} \node[var, erased] at (-0.55,{(\i-0.5)*\s}) {};
        \node[var, erasedpale] at (-0.55,5.5*\s) {};
        \node[var, known] at (0.5*\s,-0.55) {};
        \foreach \j in {2,3,4} \node[var, erased] at ({(\j-0.5)*\s},-0.55) {};
        \node[var, known] at (4.5*\s,-0.55) {};
        \node[gennode] at (-0.55,-0.55) {};
        \node[tlav, anchor=north east,rotate=-45] at (-0.2,-1.2) {$(c_1,v_2)$};
        \draw[pbrace, trose] (-0.9,1*\s) -- (-0.9,5*\s) node[midway, left=5pt, trose] {$\sub_V$};
        \draw[pbrace, trose] (4*\s,-0.9) -- (1*\s,-0.9) node[midway, below=5pt, trose] {$\sub_C$};
        \node[tpeach] at (2.5*\s,3*\s) {$\pqnbhd(\sub)$};
    \end{tikzpicture}
    \caption{
        A valid pair $(\gen,\sub)$, drawn in the local view of $\gen \sim (c_1,v_2)$.
        The set $\sub = \sub_V \sqcup \sub_C$ consists of erased qubits (pink).
        Qubits of $\supp(\gen)$ that are erased but not in $\sub$ are pale pink, and qubits outside the erasure are white.
        The checks in the pair neighborhood $\pqnbhd(\sub)$ (orange) all have weight $(1,1)$: Each is adjacent to exactly one erased qubit in $\sub_V$ and one in $\sub_C$, and to no other erased qubit.
        The remaining checks (gray) are not used.
        Here $\wt{\sub} = 7 > 3 = \wt{\supp(\gen) \setminus \env}$.
    }
    \label{fig:valid-pair}
\end{figure}
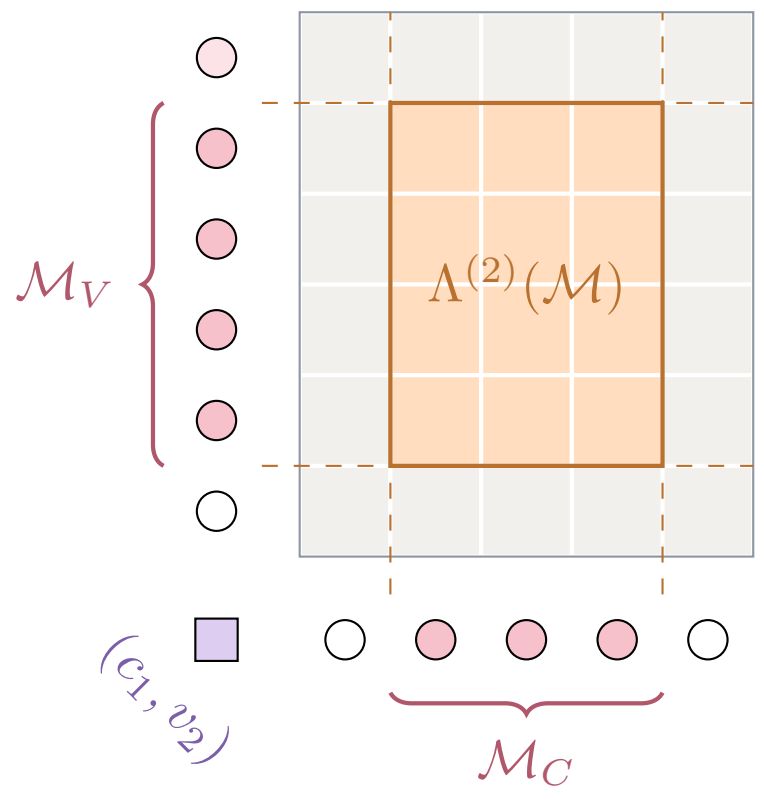

%% file: src_tikz/fill.tex
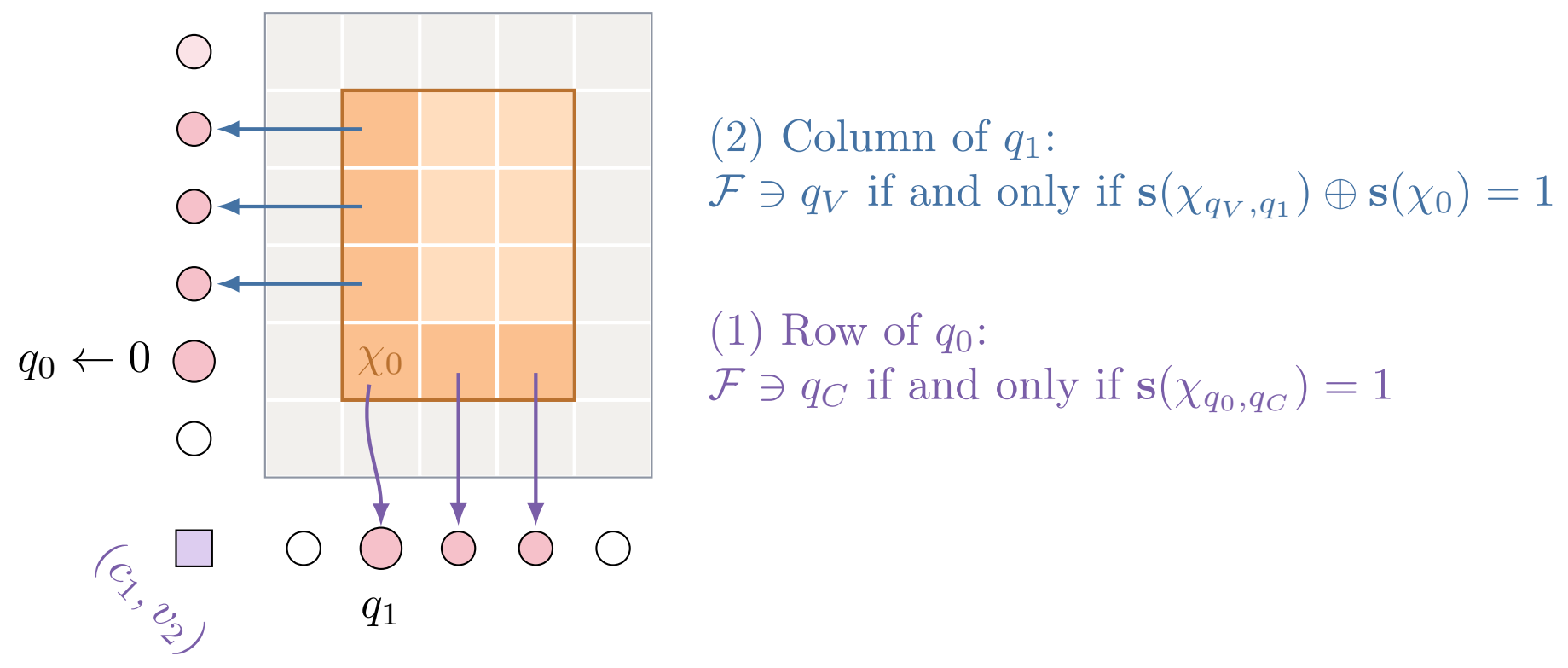
\begin{figure}[ht]
    \centering
    \begin{tikzpicture}[every node/.style={font=\small}]
        \def\s{0.6}
        \fill[pgray] (0,0) rectangle (5*\s,6*\s);
        \fill[ppeach] (1*\s,1*\s) rectangle (4*\s,5*\s);
        \fill[ppeachdeep] (1*\s,1*\s) rectangle (4*\s,2*\s);
        \fill[ppeachdeep] (1*\s,1*\s) rectangle (2*\s,5*\s);
        \draw[gridline] (0,0) grid[step=\s] (5*\s,6*\s);
        \draw[pline] (0,0) rectangle (5*\s,6*\s);
        \draw[tpeach, thick] (1*\s,1*\s) rectangle (4*\s,5*\s);
        \node[var, known] at (-0.55,0.5*\s) {};
        \node[var, erased, minimum size=3.2mm] (q0) at (-0.55,1.5*\s) {};
        \foreach \i in {3,4,5} \node[var, erased] (r\i) at (-0.55,{(\i-0.5)*\s}) {};
        \node[var, erasedpale] at (-0.55,5.5*\s) {};
        \node[var, known] at (0.5*\s,-0.55) {};
        \node[var, erased, minimum size=3.2mm] (q1) at (1.5*\s,-0.55) {};
        \foreach \j in {3,4} \node[var, erased] (c\j) at ({(\j-0.5)*\s},-0.55) {};
        \node[var, known] at (4.5*\s,-0.55) {};
        \node[gennode] at (-0.55,-0.55) {};
        \node[tlav, anchor=north east,rotate=-45] at (-0.2,-1.2) {$(c_1,v_2)$};
        \node[anchor=east] at (-0.75,1.5*\s) {$\qubit_0 \gets 0$};
        \node[anchor=north] at (1.5*\s,-0.8) {$\qubit_1$};
        \node[tpeach] at (1.5*\s,1.5*\s) {$\pc_0$};
        \foreach \j in {3,4} \draw[arrowto, tlav] ({(\j-0.5)*\s},1.35*\s) to ({(\j-0.5)*\s},-0.38);
        \draw[arrowto, tlav] (1.35*\s,1.2*\s) to[out=-100,in=90] (q1);
        \foreach \i in {3,4,5} \draw[arrowto, tblue] (1.25*\s,{(\i-0.5)*\s}) to (-0.38,{(\i-0.5)*\s});
        \node[tlav, anchor=west, align=left] at (5*\s+0.3,1.5*\s) {(1) Row of $\qubit_0$:\\ $\add \ni \qubit_C$ if and only if $\synd(\pc_{\qubit_0,\qubit_C}) = 1$};
        \node[tblue, anchor=west, align=left] at (5*\s+0.3,4*\s) {(2) Column of $\qubit_1$:\\ $\add \ni \qubit_V$ if and only if $\synd(\pc_{\qubit_V,\qubit_1}) \oplus \synd(\pc_0) = 1$};
    \end{tikzpicture}
    \caption{
        Illustrating $\Fill$ on the valid pair of Fig.~\ref{fig:valid-pair}.
        The value of $\qubit_0$, the first qubit of $\sub_V$, is fixed to $0$.
        (1) Each check in the row of $\qubit_0$ has exactly two erased qubits, $\qubit_0$ and some $\qubit_C \in \sub_C$, so its syndrome determines $\qubit_C$.
        (2) Each check in the column of $\qubit_1$, the first qubit of $\sub_C$, then determines the corresponding $\qubit_V \in \sub_V$, using the value of $\qubit_1$ found in step (1) via $\pc_0$.
    }
    \label{fig:fill}
\end{figure}

%% file: src_tikz/star-condition.tex
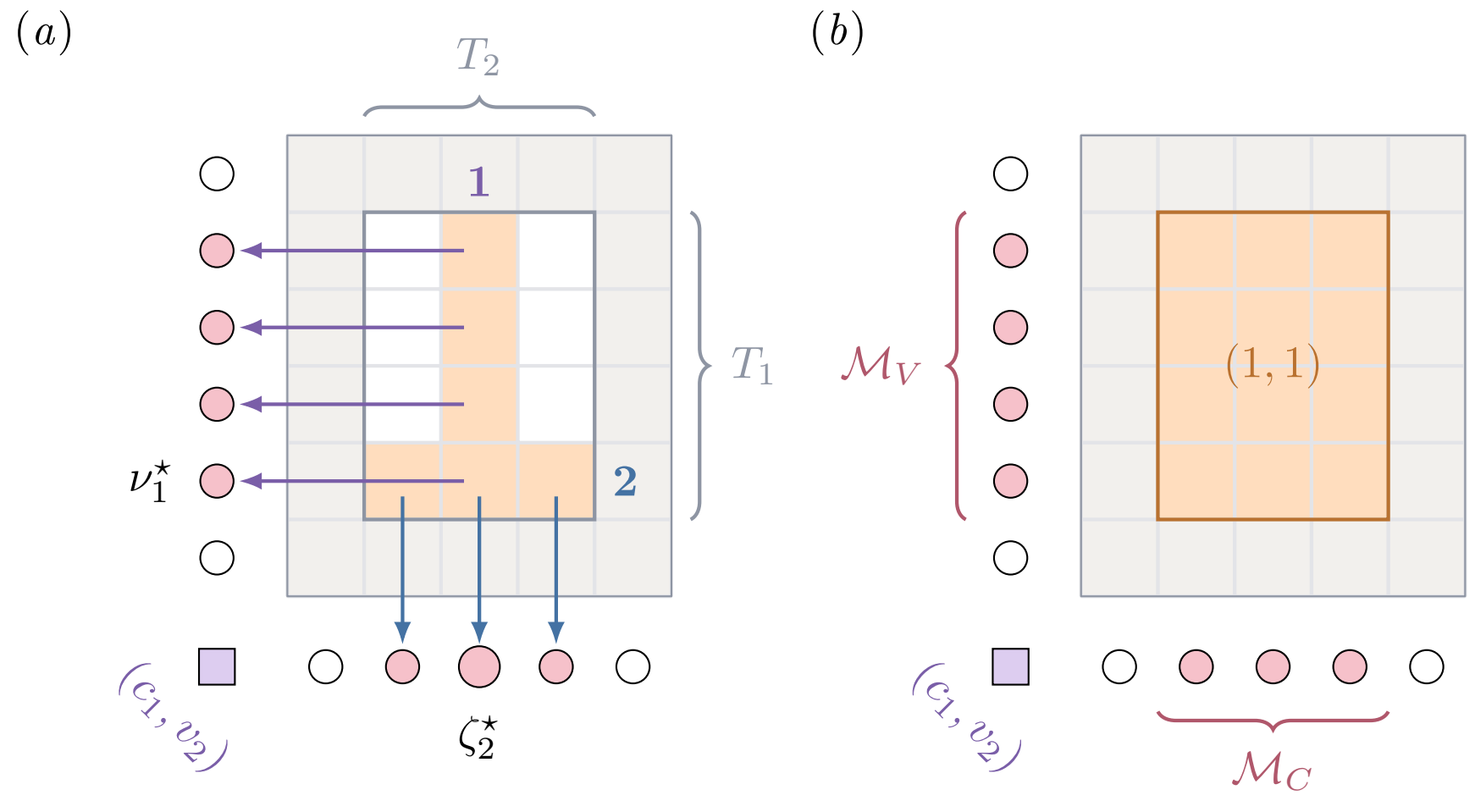
\begin{figure}[ht]
    \centering
    \begin{tikzpicture}[every node/.style={font=\small}]
        \def\s{0.6}
        \node at (-1.9,4.4) {(\emph{a})};
        \fill[pgray] (0,0) rectangle (5*\s,6*\s);
        \fill[white] (1*\s,1*\s) rectangle (4*\s,5*\s);
        \fill[ppeach] (2*\s,1*\s) rectangle (3*\s,5*\s); %
        \fill[ppeach] (1*\s,1*\s) rectangle (4*\s,2*\s); %
        \draw[gridline, draw=pline!25] (0,0) grid[step=\s] (5*\s,6*\s);
        \draw[pline] (0,0) rectangle (5*\s,6*\s);
        \draw[pline, thick] (1*\s,1*\s) rectangle (4*\s,5*\s);
        \node[var, known] at (-0.55,0.5*\s) {};
        \foreach \i in {2,...,5} \node[var, erased] at (-0.55,{(\i-0.5)*\s}) {};
        \node[var, known] at (-0.55,5.5*\s) {};
        \node[var, known] at (0.5*\s,-0.55) {};
        \node[var, erased] at (1.5*\s,-0.55) {};
        \node[var, erased, minimum size=3.2mm] at (2.5*\s,-0.55) {};
        \node[var, erased] at (3.5*\s,-0.55) {};
        \node[var, known] at (4.5*\s,-0.55) {};
        \node[gennode] at (-0.55,-0.55) {};
        \node[tlav, anchor=north east,rotate=-45] at (-0.2,-1.2) {$(c_1,v_2)$};
        \node[anchor=north] at (2.5*\s,-0.8) {$\zeta_2^\star$};
        \node[anchor=east] at (-0.75,1.5*\s) {$\nu_1^\star$};
        \foreach \i in {2,...,5} \draw[arrowto, tlav] (2.3*\s,{(\i-0.5)*\s}) -- (-0.38,{(\i-0.5)*\s});
        \node[tlav, font=\small\bfseries] at (2.5*\s,5.4*\s) {1};
        \foreach \j in {2,3,4} \draw[arrowto, tblue] ({(\j-0.5)*\s},1.3*\s) -- ({(\j-0.5)*\s},-0.38);
        \node[tblue, font=\small\bfseries] at (4.4*\s,1.5*\s) {2};
        \draw[pbrace] (5*\s+0.15,5*\s) -- (5*\s+0.15,1*\s) node[midway, right=5pt] {$T_1$};
        \draw[pbrace] (1*\s,6*\s+0.15) -- (4*\s,6*\s+0.15) node[midway, above=5pt] {$T_2$};

        \begin{scope}[xshift=6.2cm]
            \node at (-1.9,4.4) {(\emph{b})};
            \fill[pgray] (0,0) rectangle (5*\s,6*\s);
            \fill[ppeach] (1*\s,1*\s) rectangle (4*\s,5*\s);
            \draw[gridline, draw=pline!25] (0,0) grid[step=\s] (5*\s,6*\s);
            \draw[pline] (0,0) rectangle (5*\s,6*\s);
            \draw[tpeach, thick] (1*\s,1*\s) rectangle (4*\s,5*\s);
            \node[var, known] at (-0.55,0.5*\s) {};
            \foreach \i in {2,...,5} \node[var, erased] at (-0.55,{(\i-0.5)*\s}) {};
            \node[var, known] at (-0.55,5.5*\s) {};
            \node[var, known] at (0.5*\s,-0.55) {};
            \foreach \j in {2,3,4} \node[var, erased] at ({(\j-0.5)*\s},-0.55) {};
            \node[var, known] at (4.5*\s,-0.55) {};
            \node[gennode] at (-0.55,-0.55) {};
            \node[tlav, anchor=north east,rotate=-45] at (-0.2,-1.2) {$(c_1,v_2)$};
            \draw[pbrace, trose] (-0.9,1*\s) -- (-0.9,5*\s) node[midway, left=5pt, trose] {$\sub_V$};
            \draw[pbrace, trose] (4*\s,-0.9) -- (1*\s,-0.9) node[midway, below=5pt, trose] {$\sub_C$};
            \node[tpeach] at (2.5*\s,3*\s) {$(1,1)$};
        \end{scope}
    \end{tikzpicture}
    \caption{
        The $\star$-conditions in the proof of Lemma~\ref{lem:prime-existence}, in the local view of $\gen \sim (c_1,v_2)$, where $T_1 = \nbhd(c_1) \cap \unbhd(\indCone)$ indexes rows and $T_2 = \nbhd(v_2) \cap \unbhd(\envVtwo)$ indexes columns.
        Checks in the gray rows and columns are ignored.
        By Claim~\ref{claim:local-weights}, every check in $T_1 \times T_2$ can only be adjacent to erased qubits in its own row and column.
        (\emph{a}) Suppose $\unq = \emptyset$.
        (1) The qubit $(c_1,\zeta_2^\star) \in \indC$ is erased, so each check in its column would have weight $(0,1)$ unless its row qubit is erased; hence every row in $T_1$ is erased.
        (2) Each check in the row of some $\nu_1^\star \in T_1$ then forces its column qubit to be erased; hence every column in $T_2$ is erased.
        (\emph{b}) The resulting set $\sub = (T_1 \times \{v_2\}) \sqcup (\{c_1\} \times T_2)$, all of whose pair neighbors have weight $(1,1)$.
    }
    \label{fig:star-condition}
\end{figure}